\documentclass[a4paper,11pt]{article}
\usepackage{geometry}
\usepackage{a4wide}
\usepackage{graphicx}
\usepackage{amsmath}
\usepackage{amssymb}

\usepackage{cite}
\usepackage{multirow,tabularx}
\usepackage{appendix}
\usepackage{tikz}
\usetikzlibrary{shapes}
\usepackage{graphicx,amsmath,amsfonts,amssymb,amsthm,euscript,braket,xcolor}
\newcommand{\be}{\begin{equation}}
\newcommand{\ee}{\end{equation}}

\newcommand{\Rmnum}[1]{\expandafter\@slowromancap\romannumeral #1@}
\newcommand{\bea}{\begin{eqnarray}}
\newcommand{\eea}{\end{eqnarray}}

\numberwithin{equation}{section}

\usepackage{subfigure}

\begin{document}
\title{\bf Analytic three-dimensional primary hair charged black holes and thermodynamics}

\author{\textbf{Supragyan Priyadarshinee}\thanks{518ph1009@nitrkl.ac.in}, \textbf{Subhash Mahapatra}\thanks{mahapatrasub@nitrkl.ac.in}
 \\\\
 \textit{{\small Department of Physics and Astronomy, National Institute of Technology Rourkela, Rourkela - 769008, India}}
}
\date{}


\maketitle
\abstract{}
We present and discuss new families of primary hair charged black hole solutions in asymptotically anti–de Sitter space in three dimensions. The coupled Einstein-Maxwell-scalar gravity system, that carries the coupling $f(\phi)$ between the scalar and Maxwell fields is solved, and exact hairy black hole solutions are obtained analytically. The hairy solutions are obtained for three different profiles of the coupling function: (i) $f(\phi)=1$, corresponding to no direct coupling between the scalar and Maxwell fields; (ii) $f(\phi)=e^{-\phi}$; and (iii) $f(\phi)=e^{-\phi^2/2}$, corresponding to nonminimal coupling between them. For all these couplings the scalar field and curvature scalars are regular everywhere outside the horizon. We analyze the thermodynamics of the hairy black hole and find drastic changes in its thermodynamic structure due to the scalar field. For $f(\phi)=1$, there exists a critical value of the hairy parameter above which the charged hairy black hole exhibits the Hawking/Page phase transition, whereas no such phase transition occurs below this critical value. Similarly, for $f(\phi)=e^{-\phi}$ and $f(\phi)=e^{-\phi^2/2}$, the hairy solution exhibits a small/large black hole phase transition for above critical values of the hairy parameter. Interestingly, for these couplings, the thermodynamic phase diagram of three-dimensional hairy charged black holes resembles that of a higher-dimensional Reissner-Nordstr\"{o}m anti–de Sitter black hole, albeit with two second-order critical points.

\section{INTRODUCTION}
Black holes are the remarkable predictions of general relativity (GR) that play a fundamental role in our understanding of the Universe at large scales. These are the simplest and yet the most mysterious outcomes of Einstein's gravitational field equations that have been an engaging area of research for many decades and still are far from being completely understood.  They provide a unique platform where strong gravity, quantum phenomenon, and thermodynamics are believed to coexist. Although a complete understanding of the quantum mechanical nature of the black hole is still in fancy; however, great progress has been made in understanding its geometric and thermodynamic structures using the semiclassical approach. For instance, it is by now well-known that black holes not only behave like a thermal object, carrying temperature and entropy, but also can undergo phase transitions like ordinary thermodynamical systems \cite{Hawking:1975vcx,Gibbons:1976ue,Bekenstein:1973ur}. In particular, black holes in anti-de Sitter (AdS) spaces are not only thermodynamically stable, as opposed to the Schwarzschild black holes in asymptotically flat space, but also exhibit rich thermodynamic phase structure and undergo Hawking/Page (black hole to thermal-AdS) or liquid/gas (small to large black holes) like phase transitions \cite{Hawking:1982dh,Chamblin:1999tk,Chamblin:1999hg,Cvetic:1999ne,Sahay:2010tx,Sahay:2010wi,Dey:2015ytd,Mahapatra:2016dae}.

The black holes in general have been believed to follow the famous no-hair theorem. It simply states that black holes in the asymptotically flat space are uniquely described by their mass, charge, and angular momentum \cite{Ruffini:1971bza}. Said otherwise, black holes do not support scalar field hair outside their horizon in asymptotically flat space. The prime argument supporting the no-hair theorem is based on the strong absorbing nature of the horizon. A straightforward proof and critical remarks in favor of the no-scalar hair theorem for asymptotic flat spacetime were discussed in \cite{Bekenstein:1971hc}; see also \cite{Bekenstein:1971hc,Sudarsky:1995zg,Heusler:1992ss}. For a review on the issue of scalar hair in asymptotic flat spaces, see \cite{Herdeiro:2015waa}. Although the initial black hole no-hair theorem has been backed by several other works, see, for instance, \cite{Israel:1967wq,Wald:1971iw,Carter:1971zc,Robinson:1975bv,Mazur:1982db,Mazur:1984wz,Teitelboim:1972qx,Volkov:1990sva}, it is not a theorem in the strict mathematical sense. There are by now considerable counterexamples, such as Einstein–Yang–Mills theory \cite{Bizon:1990sr,Volkov:1989fi,Greene:1992fw}, dilatonic black holes \cite{Kanti:1995vq}, black hole with Skyrme hairs \cite{Luckock:1986tr,Droz:1991cx}, black hole hair with tensor vacuum \cite{Ovalle:2020kpd,Mahapatra:2022xea}  etc., that defy the no-hair theorem.

There are essentially two main requirements for a physically acceptable scalar hairy solution: (i) the scalar field should be regular and well-behaved everywhere outside the horizon and should fall off sufficiently fast at the asymptotic boundary; and (ii) the hairy geometry should be smooth, i.e., it should not contain any additional singularity. Additionally, the dynamical stability of hairy black hole solution under perturbations is desirable as well. There have been many attempts to endow black holes with the scalar hair. The first attempts were carried out in \cite{Bocharova,Bekenstein:1974sf,Bekenstein:1975ts,Bronnikov:1978mx} for asymptotic flat spaces. The resulting geometry resembles the extremal Reissner-Nordstr\"{o}m (RN) one but the scalar field diverges at the horizon. This solution was further shown to be unstable against scalar perturbations \cite{Bronnikov:1978mx}.  It was soon envisaged that the addition of a new gravitational scale, such as the cosmological constant, can provide an effective potential outside the black hole that may stabilize the scalar field, and therefore may give a mechanism to make the scalar field regular near the horizon. This leads to studying hairy solutions in asymptotically AdS or dS spaces. A hairy solution for the minimally coupled scalar-gravity system in the asymptotic dS space was obtained in \cite{Zloshchastiev:2004ny}; however, it turned out to be unstable \cite{Torii:1998ir}. In the case of asymptotically AdS space, stable hairy black hole solutions were found and discussed in \cite{Torii:1998ir,Torii:2001pg,Winstanley:2002jt,Martinez:2004nb,Martinez:2006an,Hertog:2004dr,Henneaux:2004zi,Henneaux:2006hk,Amsel:2006uf}. In recent years, many works addressing various physical properties of the hairy black holes in different asymptotic spaces have appeared; for a necessarily biased selection, see \cite{Herdeiro:2014goa,Herdeiro:2018wub,Hertog:2006rr,Kolyvaris:2010yyf,Gonzalez:2013aca,Dias:2011tj,
Bhattacharyya:2010yg,Dias:2011at,Anabalon:2012ih,Kleihaus:2013tba,Kolyvaris:2011fk,Kolyvaris:2013zfa,Charmousis:2009cm,Ballon-Bayona:2020xls,Guo:2021ere,Erices:2021uyu,Anabalon:2013qua,Astefanesei:2019ehu,
Priyadarshinee:2021rch,Mahapatra:2020wym}.

The study of the no-scalar hair theorem and fusion of scalar-gravity systems are not just of theoretical interest. The progress in gravitational wave detection \cite{LIGOScientific:2016aoc} and black hole shadow \cite{EventHorizonTelescope:2019dse} in recent years has provided us with an opportunity to test the no-hair theorem and constraint alternative scalar-gravity theories. In particular, these experimental observations are sensitive to the spacetime region around the horizon and might provide important information on additional matter fields around it. See \cite{Sadeghian:2011ub,Khodadi:2020jij,Vagnozzi:2022moj,Khodadi:2021gbc,Visinelli:2018utg,Bambi:2019tjh} for a discussion along this line, and \cite{Cardoso:2016ryw} for a review of experimental tests of the no-hair theorem. In addition, scalar fields also play a fundamental role in early Universe cosmology and particle physics \cite{Svrcek:2006yi,Peebles:1998qn}, and are often invoked to describe the dark matter and energy \cite{McDonald:1993ex,Linde:1982uu}. They also appear naturally in fundamental theories, such as string theory or high-energy unification theories \cite{Arvanitaki:2009fg}. Similarly, scalar fields in AdS spaces play an important role in probing strongly coupled field theory using the gauge/gravity duality. Prominent examples include QCD, quantum liquids, nonconformal plasmas, superfluidity and superconductivity etc., see \cite{Hartnoll:2009sz,Casalderrey-Solana:2011dxg} for review and references therein.

Lower-dimensional models of gravity are of considerable interest due to their simplicity over four and higher-dimensional models of gravity. They continue to be a source of great discussion for theorists, mainly because of the possible insight into quantum gravity that it offers. In the last few decades, three-dimensional BTZ (Banados-Teitelboim-Zanelli) black holes have drawn a lot of attention as simplified models for investigating conceptual issues relating to the black hole \cite{Banados:1992wn,Banados:1992gq}. In particular, GR becomes a topological field theory in three dimensions, whose dynamics can be described holographically by a two-dimensional conformal field theory living at the boundary of spacetime \cite{Brown:1986nw}. The BTZ black holes therefore provide a natural arena to test the deep and far-reaching ideas of AdS/CFT duality \cite{Maldacena:1997re}. For instance, the symmetric algebra and conserved charges associated with the boundary can be used to derive the BTZ black hole entropy \cite{Strominger:1997eq}. In addition, since three-dimensional gravity models can be formulated as a Chern-Simon (CS) theory, they have become quintessential for investigating general properties of gravity, and in particular its relationship with gauge field theories, in any spacetime dimensions \cite{Achucarro:1986uwr,Witten:1988hc}. Indeed, in spite of various fundamental differences with its higher-dimensional counterparts -- such as being devoid of any curvature singularity and being locally equivalent to pure $AdS_3$ -- the BTZ black hole does exhibit many of their characteristic features, such as the presence of event and Cauchy horizons, or their holographic and thermodynamic interpretations \cite{Carlip:1995qv}. For these reasons, the three-dimensional black holes continue to play a prominent role in aiming to enhance our understanding of gravity and of general features of gravitational interaction.

After the seminal discovery of the BTZ solution, the catalog of three-dimensional black hole solutions has expanded in different directions. This includes three-dimensional gravity with matter sources, such as the usual Maxwell source \cite{Martinez:1999qi}, higher curvature terms \cite{Bergshoeff:2009hq},  higher-rank tensor fields \cite{Perez:2013xi}, and gravitational Chern-Simons terms \cite{Ammon:2011nk}. Unlike its higher-dimensional counterparts, a finite number of higher-rank tensor fields can be included in three dimensions \cite{Perez:2013xi}. Similarly, after the realization that the BTZ black hole can be dressed with the regular scalar field \cite{Martinez:1996gn,Henneaux:2002wm}, several hairy black holes with minimally or nonminimally coupled self-interacting real scalar field in three spacetime dimensions are presented \cite{Chan:1994qa,Chan:1996rd,Ayon-Beato:2004nzi,Banados:2005hm,Correa:2010hf,Correa:2012rc,Xu:2013nia,Xu:2014uha,Cardenas:2014kaa,Tang:2019jkn,Dehghani:2017thu,Dehghani:2018qvn,Dehghani:2017zkm,Bueno:2021krl,Ahn:2015uza,
Karakasis:2022fep,Erices:2017izj,Zou:2014gla,Sadeghi:2013gmf,Zhao:2013isa,Bravo-Gaete:2014haa,Baake:2020tgk,Bravo-Gaete:2020ftn,Karakasis:2023ljt}. These hairy solutions include conformally coupled scalar fields, along with (or lack of) different types of coupling between the scalar and gauge fields. In many cases, these three-dimensional gravity models are also analytically tractable and have attracted considerable attention, especially for their application in holography and to probe the physics of condensed matter systems. Unfortunately, not all of these hairy solutions are physically desirable as some of them have a logarithmic radial dependent profile for the scalar field \cite{Chan:1994qa,Karakasis:2022fep}. The above-mentioned works express only a small part of the substantial attention that the three-dimensional scalar hairy black holes have gathered in recent years (see for instance \cite{Karakasis:2022fep} and references therein).

From the thermodynamic point of view, BTZ black hole behaves very differently from its higher-dimensional counterpart. In particular, the black hole entropy has multiple branches as a function of temperature in four and higher dimensions and hence can undergo phase transitions as the temperature is varied. The Hawking/Page phase transition between AdS black hole and thermal-AdS in the grand canonical ensemble and the liquid/gas type phase transition between the small and large black holes in the fixed charge ensemble are prime examples \cite{Hawking:1982dh,Chamblin:1999tk}. The latter phase transition also exhibits a second-order critical point, with critical exponents identified with the mean-field type.  However, for the BTZ black hole, for both charged and uncharged cases, the entropy vs temperature profile has only one branch and therefore does not exhibit such phase transitions. The situation also does not change much by adding scalar fields and interactions in BTZ backgrounds. See for instance \cite{Karakasis:2022fep}, where no such phase transitions in a variety of static Einstein-Maxwell-scalar (EMS) three-dimensional gravity models were found. Although certain three-dimensional EMS theories do exhibit Hawking/Page phase transition \cite{Dehghani:2017zkm}, however, the corresponding solution not only contains a logarithmic divergent profile for the scalar field but also the geometry does not asymptote to AdS at the boundary.

Since the coupling of the scalar field to the three-dimensional Einstein-Maxwell gravity generally leads to new and interesting properties for the black hole solutions, it is instructive to find new exact solutions of EMS gravity system for arbitrary coupling constants and analyze how the thermodynamical structure of black holes is modified in the presence of a scalar field. In particular, it is interesting and desirable to have three-dimensional hairy black hole solutions, with not just a regular profile of the scalar field but also whose thermodynamic structure is analogous to higher dimensional AdS black holes, as they can have many applications in applied holography. In this work, we find a family of such solutions.

In this work, we present novel analytic stable hairy charged BTZ-like solutions in the EMS theory in three dimensions, whose thermodynamic structure resembles to charged AdS black holes in higher dimensions. In particular, we consider EMS theory, with coupling function $f(\phi)$ between the scalar and Maxwell fields, and solve the coupled Einstein-Maxwell-scalar field equations simultaneously in terms of functions $A(z)$ and coupling $f(\phi)$ (see the next section for details) using the potential reconstruction method \cite{Priyadarshinee:2021rch,Mahapatra:2020wym,Dudal:2017max,Dudal:2021jav,Bohra:2020qom,Mahapatra:2018gig,Bohra:2019ebj,He:2013qq,Arefeva:2018hyo,Arefeva:2020byn,Alanen:2009xs,Arefeva:2022avn,Arefeva:2020vae,Dudal:2018ztm}. In this method, the self-interacting scalar potential form is not fixed from the beginning but is determined by the consistency of the field equations.  This method is pursued because the choice of different reasonable potentials led to a system of equations which are difficult to solve. Accordingly, as is usually done in the literature, we considered a bottom-up approach and found the gravity theory in which the field equations can be solved analytically determining also the scalar potential.

The different forms of $A(z)$ and $f(\phi)$ then allow us to construct a different family of hairy black hole solutions.  To make the analysis and results more comprehensive, we choose two particular forms of $A(z)=-\log{(1+a^2 z^2)}$ and $A(z)=-a^2 z^2$, that allow us to introduce a parameter $a$, which controls the strength of the scalar field or hair.  Similarly, we consider three different forms of coupling functions:  (i) $f(\phi)=1$; (ii)  $f(\phi)=e^{-\phi}$; and (iii) $f(\phi)=e^{-\phi^2/2}$. The first coupling function corresponds to no direct coupling between the scalar and gauge fields, whereas the second and third coupling functions correspond to nonminimal coupling between them. The prime motivation for considering such coupling functions is that they have been thoroughly investigated in various hairy black hole contexts in higher dimensions in recent years, from scalarization to holographic model building, and have constantly contributed to our ever improving understanding of the hairy aspects of black holes \cite{Herdeiro:2018wub}. It is, therefore, interesting to investigate how these coupling functions leave their imprints on the geometric and thermodynamic properties of the three-dimensional black holes as well.

We find that for these forms of $A(z)$ and $f(\phi)$, the obtained hairy gravity solution displays many desirable properties such as (i) the scalar field is regular and finite everywhere outside the horizon and falls off at the asymptotic boundary; (ii) the Kretschmann and Ricci scalars are finite everywhere outside the horizon, suggesting no additional singularity in the solution; (iii) these solutions can be analytically continued to standard BTZ solution in the limit $a\rightarrow 0$; and (iv) the scalar potential is bounded from above from its UV boundary value, thereby satisfying the Gubser criterion to have  a well-defined boundary theory \cite{Gubser:2000nd}, and reduces to the negative cosmological constant at the asymptotic boundary.

We then investigate the thermodynamics of the constructed hairy black hole solution and find that the thermodynamic behavior of the black hole becomes completely different when the hairy parameter $a$ is switched on. In particular, for $f(\phi)=1$ and a fixed charge $q_e$, there exists a critical value of the hairy parameter $a=a_c$ above which the charged hairy black hole exhibits the Hawking/Page phase transition, whereas below $a_c$ no such phase transition occurs. The corresponding transition temperature is found to be increasing (decreasing) monotonically with $a$ ($q_e$).  The critical value $a_c$ also turns out to be a $q_e$ dependent quantity, i.e., its magnitude increases as $q_e$ increases. The thermodynamic structure of the hairy black hole is therefore akin to the BTZ black hole for $a<a_c$ whereas it resembles RN-AdS black hole in the grand canonical ensemble for $a>a_c$. The thermodynamic structure of the hairy black hole becomes even more interesting for $f(\phi)=e^{-\phi}$ and $f(\phi)=e^{-\phi^2/2}$ couplings. For a fixed $q_e$, now small/large black hole phase transition appears for higher values of $a$ whereas it ceases to exist at smaller values of $a$. Interestingly, there are now two second-order critical points $\{q_{e}^{c}, a_c\}$ at which the first-order small/large black hole phase transition line stops. The small/large phase transition temperature is also found to be decreasing with $q_e$. This behavior is completely analogous to the thermodynamic behavior of the charged RN-AdS black holes in the canonical ensemble in four and higher dimensions. We moreover analyze the thermodynamic structure by varying the parameter $a$ and, interestingly, find it to be different from the varying charge scenario. In particular, now not only did the smaller values of $a$ lead to the dissolution of the small/large black hole phase transition but also the transition temperature increases with $a$. To the best of our knowledge, this is the first time a small/large black hole phase transition is observed in three dimensions.

At this point, we would like to emphasize that the constructed hairy solutions here correspond to the primary hair. In particular, one should distinguish primary hair from secondary hair \cite{Coleman:1991ku}. The secondary hair refers to black hole structures which exist solely as the result of (well-known) primary hair such as gauge charges and hence are not really new characteristics, i.e., a primary hair endows a black hole with a new independent parameter (or the quantum number) whereas the secondary hair does not \cite{Greene:1992fw}. The charged dilaton black hole solutions \cite{Gibbons:1987ps,Garfinkle:1990qj} are examples of secondary hair, where nontrivial scalar field configuration is sourced by the electric charge (primary hair). The existence of secondary hair, therefore, does not really violate the no-hair theorem. In our work, the scalar hair is not sourced by the gauge field and can be continuously tuned to get the BTZ or charged BTZ solution in the limit $a\rightarrow 0$. Most works in the hairy black hole context are concerned with secondary hair and solutions with primary hair are rare. See \cite{Gonzalez:2013aca,Anabalon:2012sn,Charmousis:2014zaa,Coleman:1991ku}, for a few other cases where black hole solutions with primary hair were discussed.

The paper is organized as follows. We introduce the Einstein-Maxwell-scalar gravity model in three dimensions and present its analytic solution in terms of two functions $f(\phi)$ and $A(z)$ in Section 2. In Section 3, using a particular form of $A(z)$, we study the geometrical and thermodynamical properties of hairy black hole solution for the coupling $f(\phi)=1$. We repeat the calculations of Section 3 with different coupling functions $f(\phi)=e^{-\phi}$ and $f(\phi)=e^{-\phi^2/2}$ in Sections, 4 and 5 respectively. In section 6, we compute the conserved mass of the black hole and establish the primary nature of the scalar hair. Finally, some concluding remarks are given in Section 7.

\section{EINSTEIN-MAXWELL-SCALAR GRAVITY SYSTEM}
To describe three-dimensional hairy charged black holes, we start with the following most general Einstein-Maxwell-scalar gravity action
\begin{eqnarray}
&&S_{EMS} =  -\frac{1}{2 \kappa^2} \int \mathrm{d^3}x \sqrt{-g} \ \ \bigl[R-\frac{f(\phi)}{4}F_{MN}F^{MN} -\frac{1}{2}\partial_{M}\phi \partial^{M}\phi -V(\phi)\bigr] \,,
\label{actionEF}
\end{eqnarray}
where $R$ is the Ricci scalar of the three-dimensional manifold $\mathcal{M}$, $F_{MN}$ is the electromagnetic field strength tensor of the $U(1)$ gauge field $B_M$,  $\phi$ is the scalar field, and $V(\phi)$ is the potential of the scalar field $\phi$. The function $f(\phi)$ represents the coupling between scalar and $U(1)$ gauge fields. To make our analysis more complete we will mainly concentrate on three different coupling functions in this work: (i) $f(\phi)=1$, corresponding to no direct coupling between the scalar and gauge fields; (ii) $f(\phi)=e^{-\phi}$; and (iii) $f(\phi)=e^{-\phi^2/2}$; corresponding to nonminimal coupling between them. The constant $\kappa$ is related to the three-dimensional Newton constant $\kappa^2=8\pi G_3$.

The variation of the above action gives the following Einstein, Maxwell, and scalar equations:
\begin{eqnarray}
& & R_{MN}- \frac{1}{2}g_{MN} R + \frac{f(\phi)}{4}\biggl(\frac{g_{MN}}{2} F^2 - 2 F_{MP}F_{N}^{\ P}\biggr)    \nonumber \\
& &  + \frac{1}{2} \biggl(\frac{g_{MN}}{2} \partial_{P}\phi \partial^{P}\phi -\partial_{M}\phi \partial_{N}\phi  + g_{MN} V(\phi)  \biggr)  =0 \,,
\label{EinsteinEE}
\end{eqnarray}
\begin{eqnarray}
& & \nabla_{M} \biggl[ f(\phi) F^{MN}  \biggr] = 0\,,
\label{MaxwellEE}
\end{eqnarray}
\begin{eqnarray}
& & \frac{1}{\sqrt{-g}}\partial_{M} \biggl[ \sqrt{-g}  \partial^{M}\phi \biggr] - \frac{F^2}{4} \frac{\partial f(\phi)}{\partial \phi} - \frac{\partial V(\phi)}{\partial \phi} = 0 \,.
\label{dilatonEE}
\end{eqnarray}
Since we want to construct static hairy black hole solutions in three dimensions, we consider the following  \textit{Ans\"atze} for the metric $g_{MN}$, gauge field $B_M$, and scalar field $\phi$:
\begin{eqnarray}
& & ds^2=\frac{L^2}{z^2}\biggl[-g(z)dt^2 + \frac{ e^{2A(z)} dz^2}{g(z)} + d\varphi^2 \biggr]\,, \nonumber \\
& & \phi=\phi(z), \ \ B_{M}=B_{t}(z)\delta_{M}^{t} \,,
\label{ansatz}
\end{eqnarray}
where $A(z)$ is the form factor, whose form will play an important role in the thermodynamics of the hairy black hole. $g(z)$ is the blackening function, and $L$ is the AdS length scale, which will be set to
one for simplicity. The radial coordinate $z$ runs from $z=0$ (asymptotic boundary) to $z=z_h$ (horizon radius), or to $z=\infty$ for thermal-AdS (without horizon).

Plugging the above \textit{Ans\"atze} into Eqs.(\ref{EinsteinEE}), we get the following three Einstein equations of motion,
\begin{eqnarray}
& & tt: \frac{A'(z)}{2 z}+\frac{L^2 e^{2 A(z)} V(\phi)}{4 z^2 g(z)}+\frac{z^2 B_{t}'(z)^2 f(\phi
   )}{8 L^2 g(z)}-\frac{g'(z)}{4 z g(z)}+\frac{1}{2 z^2}+\frac{1}{8} \phi '(z)^2 =0\,,
\label{Einsteintt}
\end{eqnarray}
\begin{eqnarray}
& & zz:  \frac{L^2 e^{2 A(z)} V(\phi)}{2 z^2}+\frac{z^2 B_{t}'(z)^2 f(\phi)}{4
   L^2}-\frac{g'(z)}{2 z}+g(z) \left(\frac{1}{z^2}-\frac{1}{4} \phi '(z)^2\right) =0\,,
\label{Einsteinzz}
\end{eqnarray}
\begin{eqnarray}
& & \varphi\varphi: g''(z) + g'(z) \left(-A'(z)-\frac{2}{z}\right) +g(z) \left(\frac{2 A'(z)}{z}+\frac{2}{z^2}+\frac{1}{2}
   \phi '(z)^2\right)\nonumber \\
& & +\frac{L^2 e^{2 A(z)} V(\phi)}{z^2}-\frac{z^2 B_{t}'(z)^2 f(\phi
   )}{2 L^2} =0\,,
\label{Einsteiny1y1}
\end{eqnarray}
where the prime denotes the derivative with respect to $z$. These complicated-looking Einstein equations can be further rearranged in the following simpler forms, which are then much
easier to analyze:
\begin{eqnarray}
& & \frac{A'(z)}{2 z}+\frac{1}{4} \phi '(z)^2 =0\,,
\label{phieom}
\end{eqnarray}
\begin{eqnarray}
& & g''(z) + g'(z) \left(-A'(z)-\frac{1}{z}\right) -\frac{z^2 f(\phi) B_{t}'(z)^2}{L^2}=0\,,
\label{geom}
\end{eqnarray}
\begin{eqnarray}
A'(z) \left(\frac{1}{2 z}-\frac{g'(z)}{4 g(z)}\right)+\frac{L^2 e^{2 A(z)} V(\phi)}{2 z^2
   g(z)}+\frac{g''(z)}{4 g(z)}-\frac{3 g'(z)}{4 z g(z)}+\frac{1}{z^2}=0\,.
\label{Veom}
\end{eqnarray}
Similarly, we get the following equation of motion for the gauge field
\begin{eqnarray}
& & \left(\frac{e^{-A(z)}f(\phi) z }{L} B_{t}'(z) \right)'=0\,,
\label{Ateom}
\end{eqnarray}
and the equation of motion for the scalar field,
\begin{eqnarray}
\phi ''(z) + \phi '(z) \left(-A'(z)+\frac{g'(z)}{g(z)}-\frac{1}{z}\right)-\frac{L^2 e^{2 A(z)}}{z^2 g(z)}\frac{\partial V(\phi)}{\partial \phi}
+\frac{z^2 B_{t}'(z)^2}{2 L^2 g(z)} \frac{\partial f(\phi)}{\partial \phi} =0\,.
\label{phieom1}
\end{eqnarray}
Therefore, we have in total five equations of motion. However, it can be explicitly checked that only four of them are independent. The last Eq.~(\ref{phieom1}) follows from the Bianchi identity and is therefore redundant. Below we will choose Eq.~(\ref{phieom1}) as a constrained equation and consider the rest of the equations as independent. To solve these equations, we further impose the following boundary conditions:
\begin{eqnarray}
&& g(0)=1 \ \ \text{and} \ \ g(z_h)=0, \nonumber \\
&& A(0) = 0 \,.
\label{boundaryconditions}
\end{eqnarray}
The boundary conditions at $z=0$ are chosen to ensure that the spacetime asymptotes to AdS at the boundary $z \rightarrow 0$. Similarly, we demand that the blackening function $g(z)$ go to zero at the black hole horizon $z_h$. Apart from these boundary conditions, we further impose the requirement that the scalar field $\phi$ goes to zero at the boundary $\phi(0)=0$ and it remains real everywhere in the bulk.

Interestingly, we can find analytic solutions of the equations (\ref{phieom})-(\ref{Ateom}) in complete closed form in terms of two functions $A(z)$ and $f(\phi)$ by the following approach: \footnote{There are actually three ways to proceed and solve the coupled Eqs.~(\ref{phieom})-(\ref{Ateom}) simultaneously.  The first and obvious one is to know the potential $V$, and then compute other variables like  $g$, $\phi$, etc.,  by solving the differential equations.  The second one is to postulate the scalar field $\phi$ and get the other variables in terms of it. The third approach is to postulate the form factor $A$, generally motivated by phenomenological consideration, and get the other variables in terms in terms of $A$. This third approach, in particular, is quite straightforward to implement and is also generally adopted in AdS/QCD model building \cite{Dudal:2017max,Bohra:2019ebj}.}
\begin{itemize}
\item Solve Eq.~(\ref{Ateom}) and find $B_t(z)$ in terms of $A(z)$ and $f(\phi)$.

\item From the obtained $B_t(z)$ solution, solve Eq.~(\ref{geom}) and find $g(z)$ in terms of $A(z)$ and $f(\phi)$.

\item Solve Eq.~(\ref{phieom}) and find $\phi'(z)$ in terms of $A(z)$.

\item Lastly, solve Eq.~(\ref{Veom}) and find $V$ in terms of $A(z)$ and $g(z)$.
\end{itemize}
Using this approach, we get the solution for $B_t(z)$ from Eq.~(\ref{Ateom}) as
\begin{eqnarray}
& & B_{t}' (z) = - \frac{q_e L e^{A(z)}}{ z f(z)} \,, \nonumber\\
& & B_{t} (z) = - q_e L  \int dz \frac{e^{A(z)}}{z f(z)} \,,
\label{Atsol}
\end{eqnarray}
where the integration constant $q_e$ is related to the total charge of the black hole (see below).
Now, substituting Eq.~(\ref{Atsol}) into Eq.~(\ref{geom}), we get the following solution for $g(z)$,
\begin{eqnarray}
& & g(z) =  C_1 + \int_0^z \, d\xi \ e^{A(\xi)} \xi \biggl[ C_{2} + \mathcal{K}(\xi) \biggr] \,,
\label{gsol}
\end{eqnarray}
with,
\begin{eqnarray}
& & \mathcal{K}(\xi)= \int \, d\xi \ \biggl[\frac{ B_{t}'^2 e^{-A(\xi)} \xi f(\xi) }{ L^2 } \biggr] \,,
\label{gsol}
\end{eqnarray}
where the integration constants $C_1$ and $C_2$ [obtained from Eq.~(\ref{boundaryconditions})] are given as
\begin{eqnarray}
C_1 = 1,  \ \ \ \ \ C_2 = - \frac{1+ \int_0^{z_h} \, d\xi e^{A(\xi)} \xi \mathcal{K}(\xi) }{ \int_0^{z_h} \, d\xi e^{A(\xi)} \xi}  \,.
\end{eqnarray}
Similarly, the scalar field can be solved in terms of $A(z)$ from Eq.~(\ref{phieom})
\begin{eqnarray}
\phi(z) = \int \, dz \  \sqrt{-\frac{2 A'(z)}{z}} + C_3 \,,
\label{phisol}
\end{eqnarray}
where the constant $C_{3}$ can be fixed by demanding $\phi$ to vanishes near the asymptotic boundary, i.e., $\phi |_{z=0}\rightarrow 0$. Lastly, the potential $V$ can be found from Eq. (\ref{Veom}),
\begin{eqnarray}
V(z) = -\frac{e^{-2 A(z)}}{4 L^2} \left(-2 z \left(z A'(z)+3\right) g'(z)+4 g(z) \left(z A'(z)+2\right)+2 z^2 g''(z) \right) \,.
\label{Vsol}
\end{eqnarray}
It is thus clear that one can systematically obtain a closed-form analytic solution of the Einstein-Maxwell-scalar gravity system in $(2+1)$-dimensions in terms of two functions, i.e., $A(z)$ and $f(\phi)$, and construct hairy charged BTZ black hole solutions. Once the coupling function $f(\phi)$ is given, the constructed hairy solution will depend only on $A(z)$. The different forms of $A(z)$ and $f(\phi)$ will, however, correspond to different $V(z)$, i.e., different $A(z)$ and $f(\phi)$ will attribute to different $(2+1)$-dimensional hairy black hole solutions. Therefore, one can systematically construct a large family of physically allowed hairy charged BTZ black hole solutions for the Einstein-Maxwell-scalar gravity system in $(2+1)$-dimensions by choosing different forms of $A(z)$ and $f(\phi)$.

Nevertheless, in the context of applied gauge/gravity duality, the forms of $A(z)$ and $f(\phi)$ are usually fixed by taking guidance from the dual boundary field theory. In particular, depending
upon what type of boundary field theory one is interested in one generally considers different forms of $A(z)$ and $f(\phi)$. For instance, in the area of AdS/QCD model building, the forms of these functions are generally fixed by demanding the dual boundary field theory to display real QCD properties such as confinement/deconfinement phase transition, confinement in the quark sector, linear Regge trajectory for the excited meson mass spectrum,  etc, \cite{Dudal:2017max,Bohra:2019ebj}.

We can also take a more pragmatic and liberal route and examine various physically motivated forms of $A(z)$ and $f(\phi)$ to comprehensively discuss the effects of scalar hair and make a qualitative statement on the stability and thermodynamics of the hairy charged black holes in three dimensions, without concerning too much about the dual boundary field theory. Here, we take such a route. Particularly, we consider three different forms of the coupling function $f(\phi)$: (i) $f(\phi)=1$; (ii) $f(\phi)=e^{-\phi}$; and (iii) $f(\phi)=e^{-\phi^2/2}$. As mentioned in the introduction, these three types of couplings have been greatly considered in various hairy black hole contexts in recent years, for instance see \cite{Herdeiro:2018wub}, and hence it is interesting to analyze how these different coupling functions modify the hairy structure in three dimensions as well. Similarly, we can take different forms of $A(z)$. Following \cite{Priyadarshinee:2021rch}, we consider a particularly simple form $A(z)=-\log{(1+a^2 z^2)}$. This form of $A(z)$ is taken not just for its simplicity but also to have better control over the integrals that appear in Eqs.~(\ref{Atsol})-(\ref{Vsol}), again without concerning greatly about the dual field theory. In principle, we can choose other simple forms, such as $A(z)=-a^2 z^n$ with $n\geq1$, as well. This form of $A(z)$ has also been extensively used in the literature, see for example \cite{Mahapatra:2020wym,Dudal:2017max}. However, we will not dwell much into this form here and relegate most of our calculations to the Appendix for completeness. With the considered form of $A(z)=-\log(1+a^2 z^2)$, the strength of the scalar field is characterized by the parameter $a$. Therefore, when the parameter $a$ goes to zero so does the backreaction of the scalar field. Hence, as desired, in the limit $a\rightarrow 0$, we get back to the charged BTZ black hole solution.

There are also other important reasons for taking the above-mentioned forms of $f(\phi)$ and $A(z)$. In particular, these forms ensure that the constructed hairy geometry asymptotes to AdS at the boundary $z\rightarrow 0$, i.e., at the  boundary, we have
\begin{eqnarray}
& & V(z)|_{z\rightarrow 0} = -\frac{2}{L^2} + \frac{m^2\phi^2}{2}+\dots \,, \nonumber\\
 & &  V(z)|_{z\rightarrow 0} =  2\Lambda + \frac{m^2\phi^2}{2}+\dots   \,,
\label{Vsolexp}
\end{eqnarray}
where $\Lambda=-\frac{1}{L^2}$ is as usual the negative cosmological constant in three dimensions. Similarly, the Ricci scalar $R$ approaches $-6/L^2$ asymptotically. This, together with the fact that $g(z)|_{z\rightarrow 0}=1$, certainly establishes that the constructed spacetime asymptotes to AdS at the boundary. Moreover, $m^2=-1$ is the mass of the scalar field, satisfying the Breitenlohner-Freedman bound for stability in AdS space, i.e, $m^2\geq-1$ \cite{Breitenlohner:1982jf}. Furthermore, as we will show later on, these geometries also satisfy the Gubser criterion to have a well-defined dual boundary theory \cite{Gubser:2000nd}.

Let us also check the validity of the null energy condition (NEC) to further establish the consistency of the constructed hairy solutions. The NEC can be expressed as
\begin{eqnarray}
T_{MN}\mathcal{N}^M \mathcal{N}^N \geqslant 0 \,,
\label{NEC}
\end{eqnarray}
where $T_{MN}$ is the energy-momentum tensor of the matter fields. The null vector $\mathcal{N}^{M}$ satisfies the condition $g_{MN}\mathcal{N}^M \mathcal{N}^N=0$, and can be chosen as
\begin{eqnarray}
\mathcal{N}^M= \frac{1}{\sqrt{g(z)}}\mathcal{N}^{t} + \frac{\cos{\alpha}\sqrt{g(z)}}{e^{A(z)}} \mathcal{N}^{z} + \sin{\alpha} \mathcal{N}^{\varphi}  \,,
\label{nullvector}
\end{eqnarray}
for arbitrary parameter $\alpha$. The NEC then becomes
\begin{eqnarray}
T_{MN}\mathcal{N}^M \mathcal{N}^N = \frac{e^{-2A(z)}}{2L^2} \left(z^2 f(\phi) \sin^2{\alpha} B_{t}'(z) + L^2 \cos^2{\alpha} g(z) \phi'(z)^2   \right) \geqslant 0  \,,
\label{nullvector}
\end{eqnarray}
which is always satisfied everywhere outside the horizon for the chosen forms of $A(z)$ and $f(\phi)$.

We also like to mention that although the form of function $A(z)$ seems to be slightly arbitrary in our formalism, however, we are required to be careful while choosing its form. In particular, we need to ensure that all the fields remain real throughout the spacetime. This puts a lot of restrictions on the forms of $A(z)$ we can consider. For instance, the scalar field $\phi$ would be complex if $A(z)=a z$, with $a$ being positive, is considered. Similarly, we can not consider forms such as $A(z)=a/z^n$, with $n$ being positive, as this will make the asymptotic boundary different from AdS.

Let us now write down the expressions of various thermodynamic observables of the constructed hairy black holes. This will be useful in the discussion of black hole thermodynamics later on. The temperature ($T$) and entropy of the black hole ($S_{BH}$) are given by
\begin{eqnarray}
& & T=-\frac{e^{-A(z)}g'(z_h)}{4 \pi} = -\frac{z_{h} (C2+K(z_h))}{4 \pi} \, , \nonumber\\
& &  S_{BH}= \frac{\mathcal{A}}{4 G_3} = \frac{ 2 \pi L }{4 G_3 z_{h}} \,,
\label{STexp}
\end{eqnarray}
where $\mathcal{A}$ is the area of the event horizon. Similarly, the electric charge $Q_e$ of the black hole can be found by measuring the flux of the electric field at the boundary,
\begin{eqnarray}
Q_e=\frac{1}{16 \pi G_3} \int  f(\phi)F_{\alpha\beta}u^\alpha n^\beta \ d\varphi\,
\end{eqnarray}
where $u^\alpha$ and $n^\beta$ are the unit spacelike and timelike normals, respectively, to the constant radial surface
\begin{eqnarray}
& & u^\alpha = \frac{1}{\sqrt{-g_{tt}}} \delta^{\alpha}_{t} =\frac{z}{L \sqrt{g(z)}} \delta^{\alpha}_{t} \,, \nonumber\\
& & n^\beta = \frac{1}{\sqrt{g_{zz}}}\delta^{\beta}_{z}=\frac{z \sqrt{g(z)}}{L e^{A(z)}} \delta^{\beta}_{z} \,,
\end{eqnarray}
and $d\varphi$ represents the integration over the spatial one-dimensional boundary space. Using Eq.~(\ref{Atsol}) and simplifying, we get the following expression of charge:
\begin{eqnarray}
& & Q_e=\frac{q_e}{8 \pi G_3} \,.
\label{Qesol}
\end{eqnarray}
We can further find an explicit relation between $Q_e$ and the corresponding conjugate electric potential $\mu_e$. The black hole's electric potential on the horizon,
measured by an observer at the reference point, can be obtained in terms of the null generator of the horizon. This is given by
\begin{eqnarray}
\mu_e = B_{\alpha} \chi^\alpha|_{reference} - B_{\alpha} \chi^\alpha|_{z\rightarrow z_h} = -B_{t}(z_h)
\end{eqnarray}
where in the reference, electric potential should vanish.

Apart from the above-mentioned hairy black hole solution, there exists another allowed solution to the Einstein-Maxwell-scalar equations of motion. This solution, called thermal-AdS here, does not contain a horizon.  \footnote{For convenience, we are calling this solution as thermal-AdS here even though it does not have a constant curvature. } The thermal-AdS solution corresponds to $g(z)=1$, and can be obtained from the black hole solution by taking the limit $z_h\rightarrow \infty$. The thermal-AdS again asymptotes to AdS at the boundary, however importantly, depending on the nature of $A(z)$, it can have a nontrivial structure in the bulk. In particular, it does not have, unlike the usual AdS space in three dimensions, a constant curvature. Interestingly, as we will see later on, depending on the magnitude $a$ and $q_e$, there can also be a Hawking/Page type thermal-AdS/black hole phase transition between these two solutions.\\

\section{HAIRY BLACK HOLES WITH $f(\phi)=1$ COUPLING}
In this section, we first examine the geometry and thermodynamic properties of the charged hairy black hole solution for $f(\phi)=1$. For $f(\phi)=1$ and $A(z)=-\log{(1+a^2 z^2)}$, the solution for scalar and gauge fields reduces to
\begin{eqnarray}
& & \phi(z) = 2 \sinh ^{-1}(a z) \,, \nonumber\\
& & B_t(z) = \frac{q_e}{2} \log{\frac{1+ a^2 z^2}{z^2}}\,.
\label{phibtsolf1}
\end{eqnarray}
Notice that in the limit $a\rightarrow 0$, the scalar field vanishes and the gauge field reduces to the standard charged BTZ expression. Similarly, we have the following solution for $g(z)$,
\begin{eqnarray}
& & g(z) = 1-\frac{\log \left(1+a^2 z^2\right)}{\log \left(1+a^2 z_h^2\right)} + \frac{q_{e}^2 \log \left(1+a^2 z^2\right)\log \left(\frac{(1+a^2 z_h^2)z^4}{\left(1+a^2 z^2\right)
   z_h^4}\right)}{8 a^2} \nonumber\\
& & + q_{e}^2 \left(\frac{\text{Li}_2\left(-a^2 z^2\right)}{4 a^2}-\frac{\log \left(1+a^2 z^2\right)
   \text{Li}_2\left(-a^2 z_h^2\right)}{4 a^2 \log \left(1+a^2 z_h^2\right)}\right)\,,
\label{gsolf1}
\end{eqnarray}
where $\text{Li}_2$ is the polylogarithm function. We can check that this expression again reduces to the standard charged BTZ expression in the limit $a\rightarrow 0$. We can similarly write down the analytic expression of $V(z)$. Unfortunately, it is too lengthy, and at the same time not very illuminating; therefore, we omit to write down it here for brevity.

\begin{figure}[ht]
	\subfigure[]{
		\includegraphics[scale=0.4]{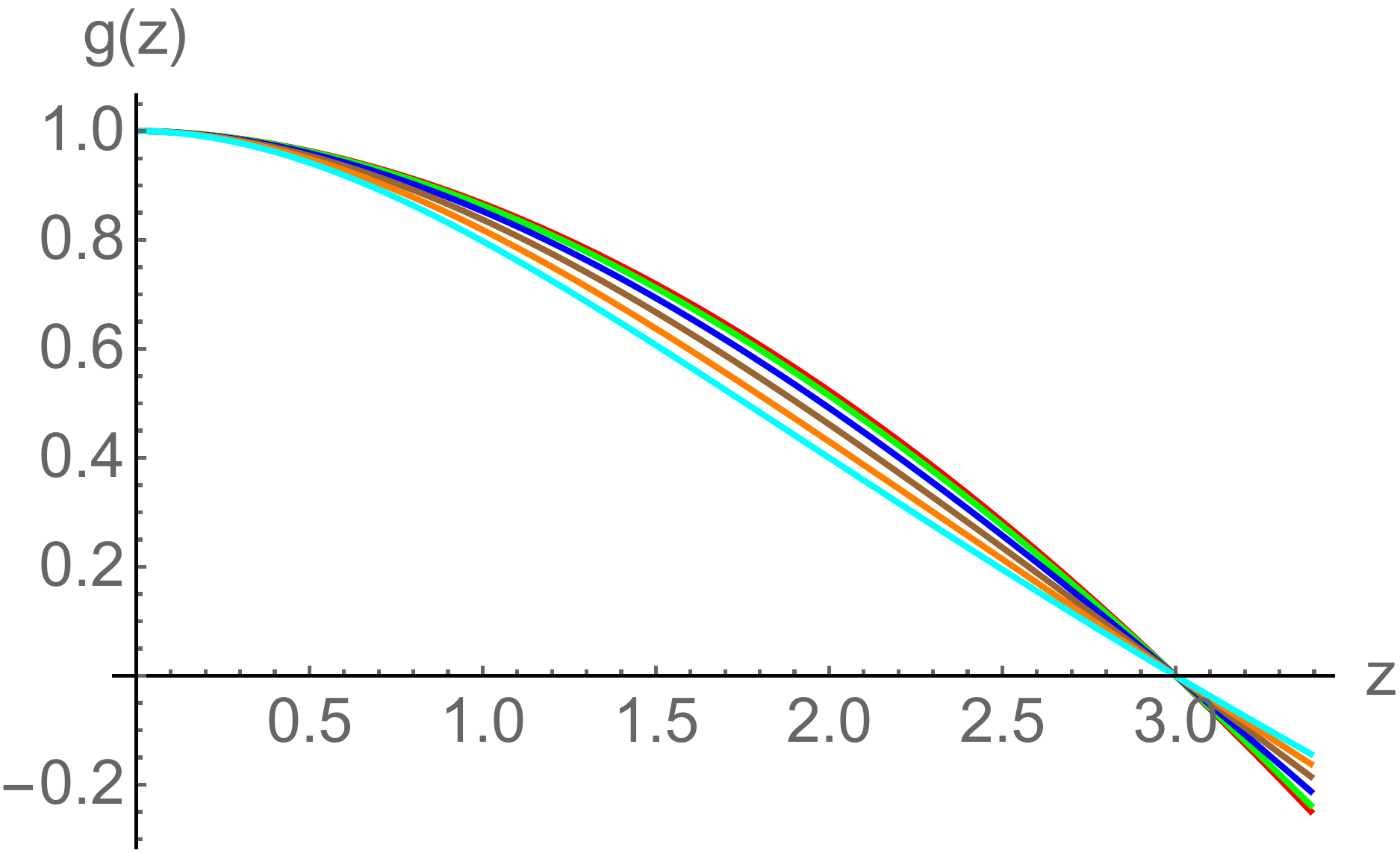}
	}
	\subfigure[]{
		\includegraphics[scale=0.4]{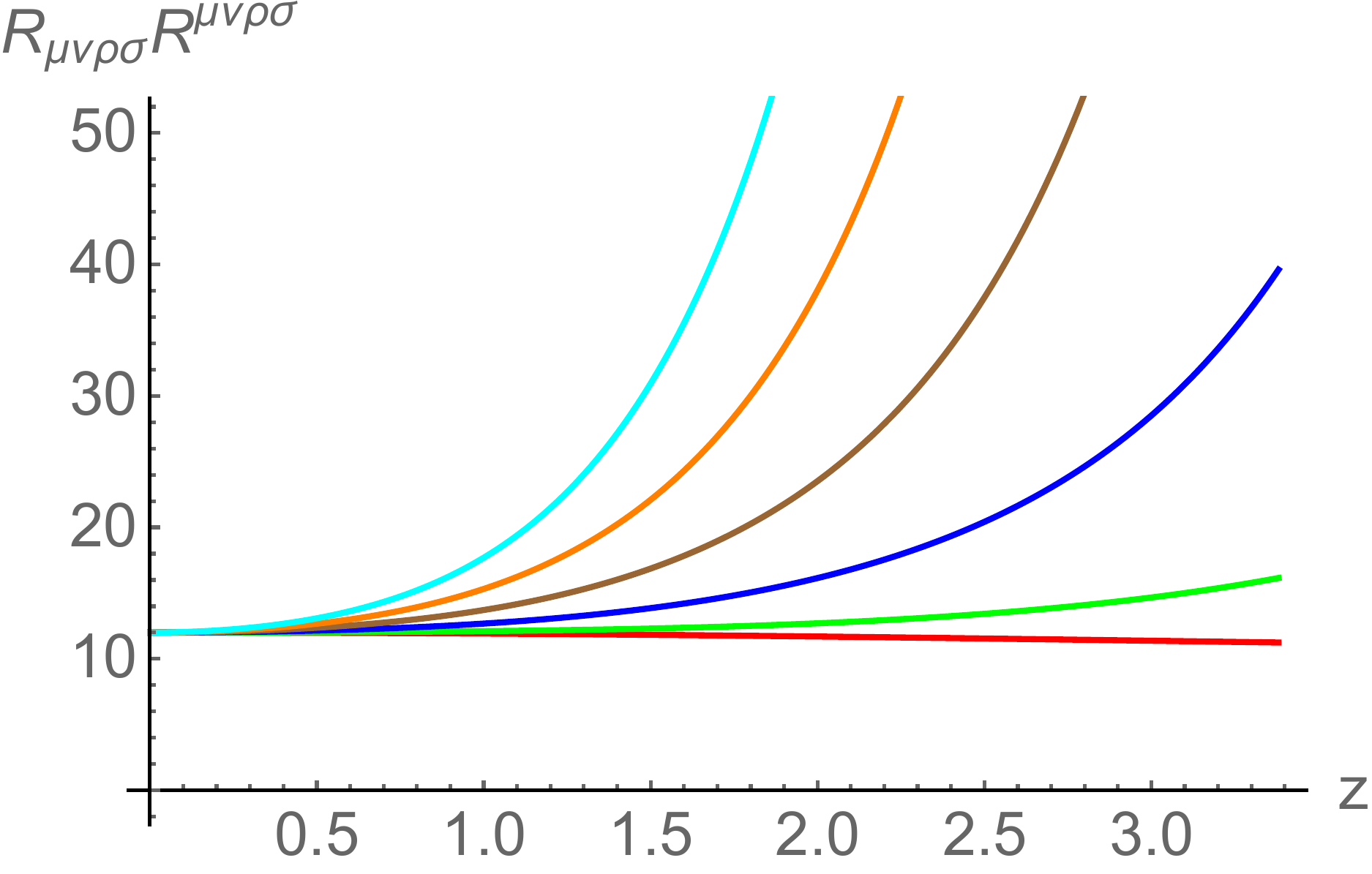}
	}
	\subfigure[]{
		\includegraphics[scale=0.4]{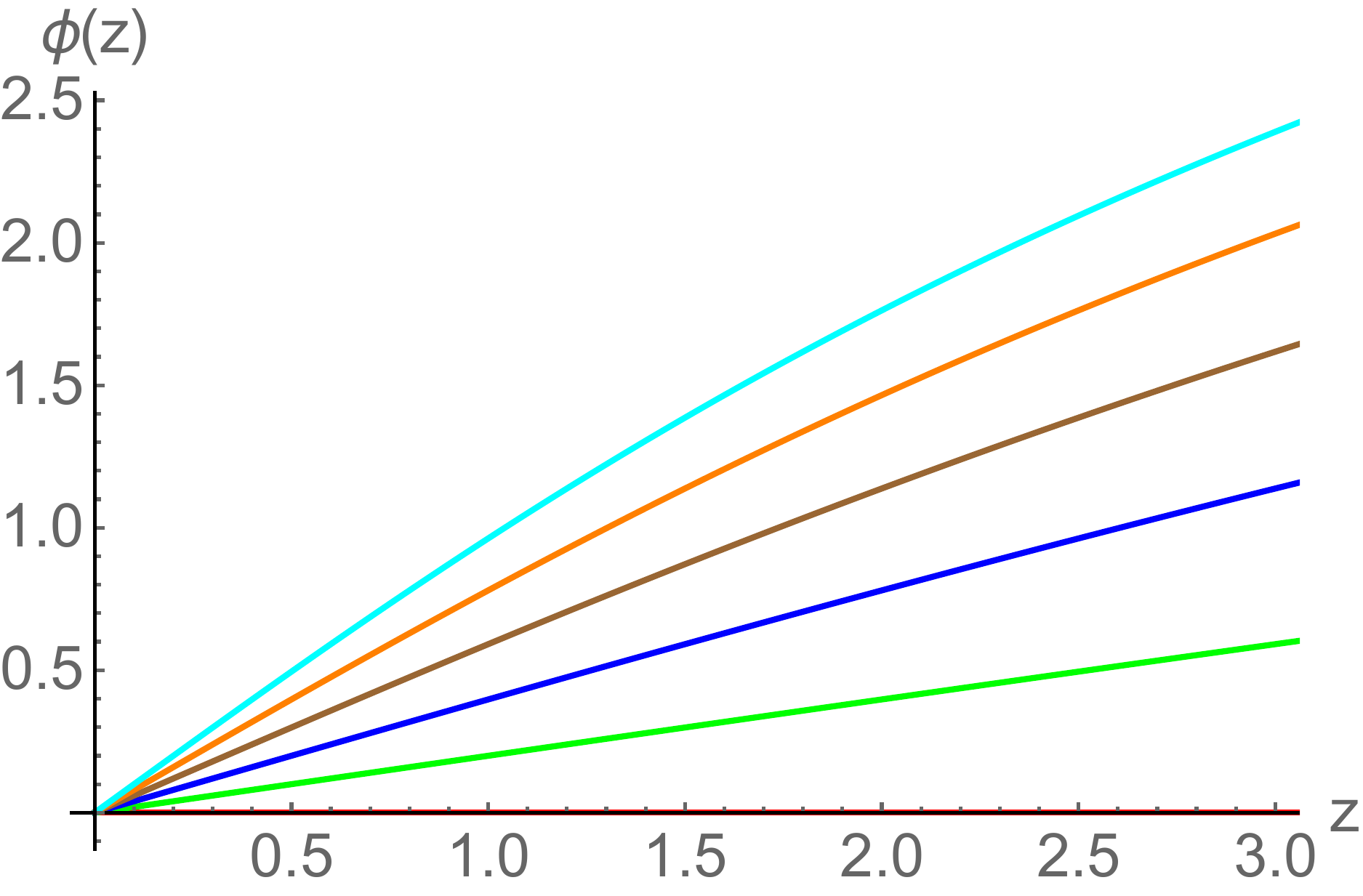}
	}
\subfigure[]{
		\includegraphics[scale=0.4]{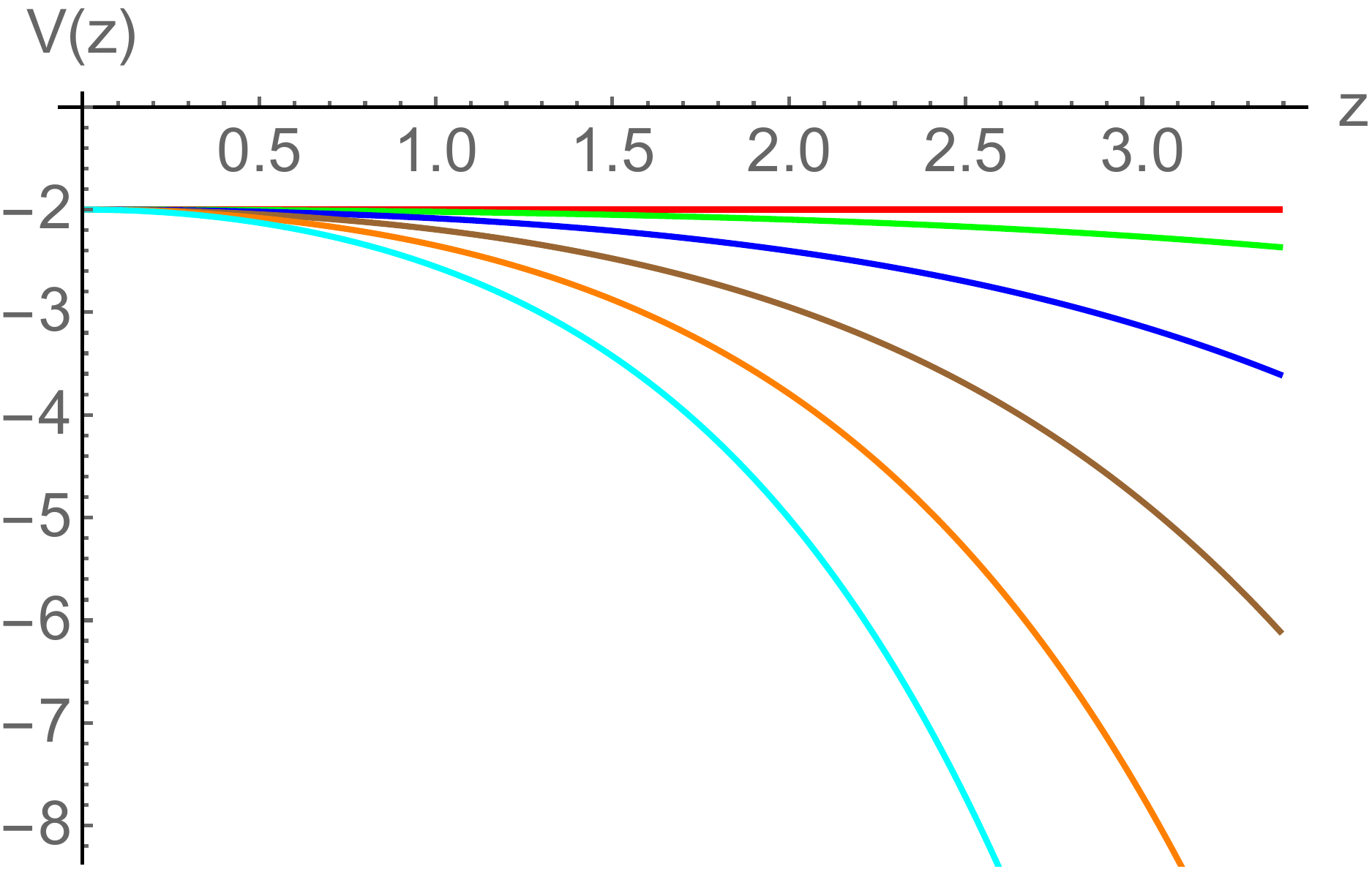}
	}
	\caption{\small The behavior of $g(z)$, $R_{MNPQ}R^{MNPQ}$, $\phi(z)$, and $V(z)$ for different values of hair parameter $a$. Here $z_h=3$ and $q_e=0.2$ are used. Red, green, blue, brown, orange, and cyan curves correspond to $a=0$, $0.1$, $0.2$, $0.3$, $0.4$, and $0.5$, respectively.}
	\label{zvsgzvsazh3qPt2f1}
\end{figure}

In Fig.~\ref{zvsgzvsazh3qPt2f1}, the radial profile of various fields for different values of scalar hair parameter $a$ is shown. The results here are shown for a particular value of $z_h=3$ and $q_e=0.2$; however, analogous results occur for their other values as well. Notice that the blackening function $g(z)$ undergoes a sign change at $z_h$, indicating the presence of a horizon. This is true for all values of $a$. Similarly, the Kretschmann scalar $R_{MNPQ}R^{MNPQ}$ is finite everywhere outside the horizon. The same is true for the Ricci scalar. This indicates the nonsingular nature of the hairy spacetime.  The curvature singularity appears only at $z=1/r=\infty$. Therefore, there is no additional singularity in the hairy black hole case than those already present at BTZ black hole. However, the strength of the singularity increases with the scalar hair. In particular, $R_{MNPQ}R^{MNPQ} \propto z^2$ for the BTZ case, whereas $R_{MNPQ}R^{MNPQ} \propto z^6 \log{z}$ for the hairy case (the singularity in the BTZ case comes from the $U(1)$ gauge field. If $q_e=0$, the
BTZ spacetime is completely regular everywhere). Accordingly, the hairy spacetime is more curved compared to its nonhairy counterpart.

Similarly, the scalar field is regular and finite everywhere outside the horizon. The scalar field is real and goes to zero only at the asymptotic boundary, suggesting the existence of a well-behaved hairy black hole solution in three dimensions.  The thermodynamic local stability of these charged hairy black hole solutions will be analyzed shortly when we will discuss their free energy and specific heat. Similarly, the potential is also finite and regular in the outside horizon region. Additionally, as desired, it asymptotes to $V(z=0)=-2$ at the boundary. This is again true for all $a$. Moreover, provided that $q_e$ is not too large, $V(z)$ is also bounded from above by its UV boundary value, i.e., $V(0)\geq V(z)$, thereby fulfilling the Gubser criterion to have a well-defined boundary theory \cite{Gubser:2000nd}. However, for larger values of charge $q_e \gtrsim 2$, the criterion can be violated.

We moreover examine the behavior of $V(z)$ with respect to $\phi(z)$. We find that the $\phi$ vs $V$ profiles nearly coincide with each other for different values of $a$ and $z_h$, confirming the almost independence of $V$ on these parameters. With the exception of large values of $q_e$, the $\phi$ vs $V(z)$ profiles for different values of $q_e$ also coincide. This is invariably a byproduct of the above-discussed violation of the Gubser criterion. Essentially, the large $q_e$ values for which the Gubser criterion is not respected also lead to unphysical $\phi$ vs $V$ behavior. In the rest of the paper, we will focus on only those parameter values for which the Gubser criterion is respected.

\begin{figure}[h!]
\begin{minipage}[b]{0.5\linewidth}
\centering
\includegraphics[width=2.8in,height=2.3in]{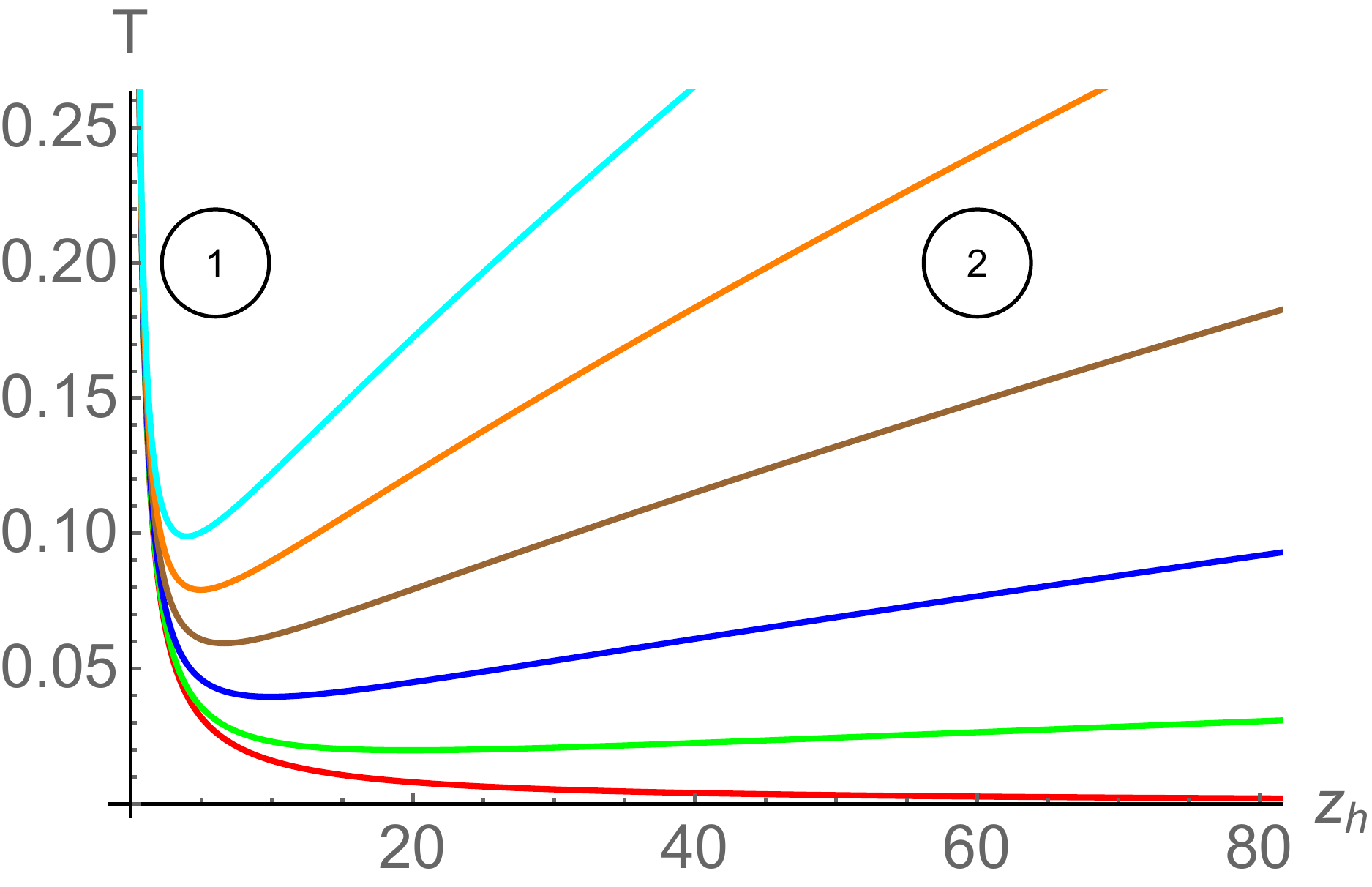}
\caption{ \small Hawking temperature $T$ as a function of horizon radius $z_h$ for various values of $a$. Here $q_e=0$ is used. Red, green, blue, brown, orange, and cyan curves correspond to $a=0$, $0.1$, $0.2$, $0.3$, $0.4$, and $0.5$, respectively.}
\label{zhvsTvsaq0f1}
\end{minipage}
\hspace{0.4cm}
\begin{minipage}[b]{0.5\linewidth}
\centering
\includegraphics[width=2.8in,height=2.3in]{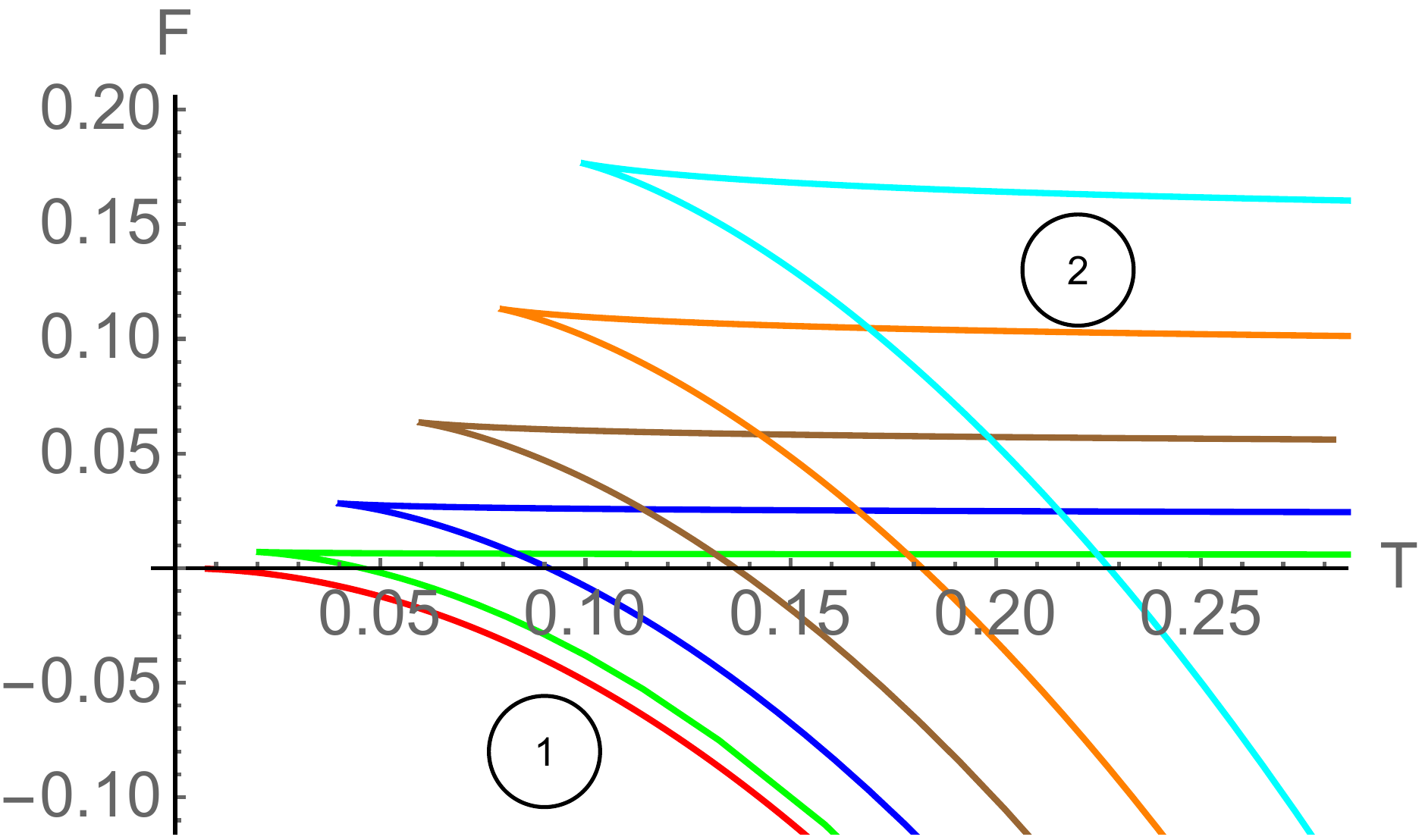}
\caption{\small Free energy $F$ as a function of Hawking temperature $T$ for various values of $a$. Here $q_e=0$ is used. Red, green, blue, brown, orange, and cyan curves correspond to $a=0$, $0.1$, $0.2$, $0.3$, $0.4$, and $0.5$, respectively.}
\label{TvsFvsaq0f1}
\end{minipage}
\end{figure}

We now move on to discuss the thermodynamics of the hairy black hole. For this purpose, let us first explicitly write down the expression of the black hole temperature. For $f(\phi)=1$, this is given by
\begin{eqnarray}
T = \frac{a^2 z_h}{2 \pi  \log \left(1+a^2 z_h^2\right)} + \frac{q_{e}^2 z_h }{8 \pi } \left( \frac{
   \log \left(1+a^2 z_h^2\right)}{2} + \frac{\text{Li}_2\left(-a^2 z_h^2\right)}{\log \left(1+a^2 z_h^2\right)}\right) \,.
\label{tempf1}
\end{eqnarray}
Note that in the limit $a\rightarrow 0$, the above expression smoothly reduces to the standard charged BTZ expression $T=(4-q_{e}^2 z_{h}^2)/(8 \pi z_h)$. In Fig.~\ref{zhvsTvsaq0f1}, the variation of temperature with respect to (inverse) horizon radius $z_h$ for different values of $a$ is shown. Here we have kept $q_e=0$ fixed, corresponding to the uncharged hairy black hole. Notice that for $a=0$ (red line) there is only one black hole branch. The temperature of this black hole branch decreases with $z_h$ and has a positive specific heat. Accordingly, this black hole branch is thermodynamically stable. The local thermodynamic stability of these black holes will be discussed shortly. This is an expected result since for $a=0$ and $q_e=0$ our hairy solution reduces to the stable uncharged BTZ black hole.

However, the thermodynamic behavior of the three-dimensional black hole becomes completely different when the hairy parameter $a$ is switched on. In particular, there are now two black hole branches: one stable and one unstable. For the stable black hole branch, the temperature decreases with $z_h$ whereas for the unstable black hole branch, the temperature increases with $z_h$ (blue line). The emergence of the second unstable black hole branch can also be analytically observed from Eq.~(\ref{tempf1}). Notice that for $q_e=0$, only the first term contributes to the temperature and this term increases with $z_h$ for large $z_h$. These stable/unstable branches are indicated by \textcircled{1} and \textcircled{2}, respectively, in Fig.~\ref{zhvsTvsaq0f1}. Interestingly, unlike the BTZ black hole, the hairy black hole branches exist only above a certain temperature. In particular, there exists a minimum temperature $T_{min}$ below which the hairy black hole ceases to exist and the thermal-AdS phase remains the only possible solution.  This is true for all finite values of $a$. Importantly, as is usually the case, the appearance of multivaluedness of the temperature also suggests a possible phase transition in the hairy black holes.

This expectation indeed turns out to be true. In Fig.~\ref{TvsFvsaq0f1}, the free energy of the hairy black holes, normalised with respect to thermal-AdS, is plotted. The color scheme used here is similar to Fig.~\ref{zhvsTvsaq0f1}. We notice that, upon varying temperature, the free energy changes its sign at some critical temperature $T_{HP}$, indicating the well-known Hawking/Page type phase transition between uncharged hairy black hole (first branch) and thermal-AdS at $T_{HP}$. In particular, above $T_{HP}$ hairy black hole is thermodynamically favored whereas below this temperature thermal-AdS is thermodynamically favored. Also, the free energy of the second black hole branch is always higher than the first black hole branch, indicating that the former is always thermodynamically disfavored with respect to the latter.

\begin{figure}[h!]
\centering
\includegraphics[width=2.8in,height=2.3in]{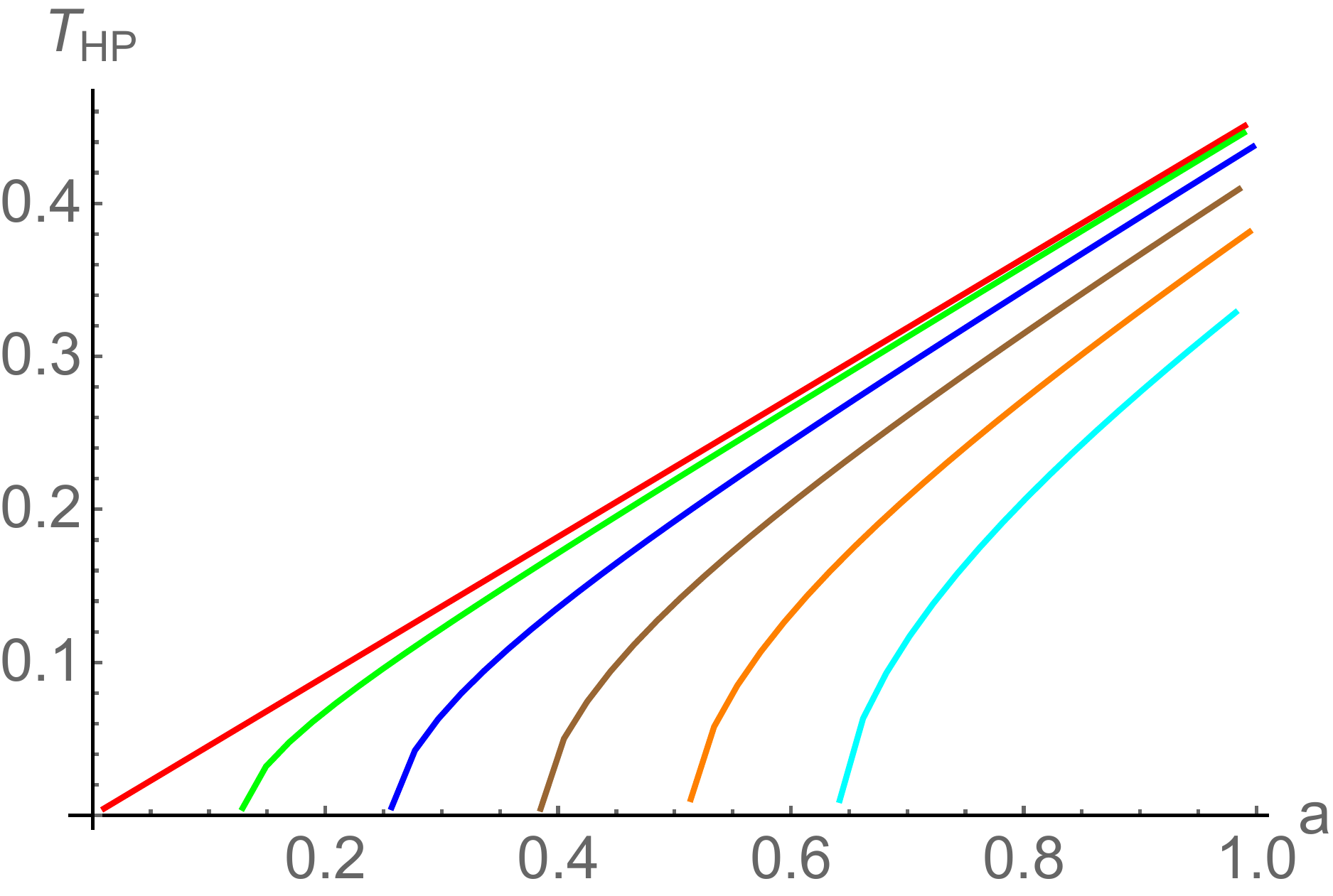}
\caption{ \small Hawking/Page phase transition temperature $T_{HP}$ as a function of $a$ for various values of $q_e$. Red, green, blue, brown, orange, and cyan curves correspond to $q_e=0$, $0.2$, $0.4$, $0.6$, $0.8$, and $1.0$, respectively.}
\label{avsTHPvsqf1}
\end{figure}

We have further analyzed the dependence of $T_{HP}$ on $a$. The complete dependence is shown in Fig.~\ref{avsTHPvsqf1}. It shows that $T_{HP}$ increases monotonically with $a$, a result that can also be readily inferred from Fig.~\ref{TvsFvsaq0f1} for the chargeless case.

\begin{figure}[h!]
\begin{minipage}[b]{0.5\linewidth}
\centering
\includegraphics[width=2.8in,height=2.3in]{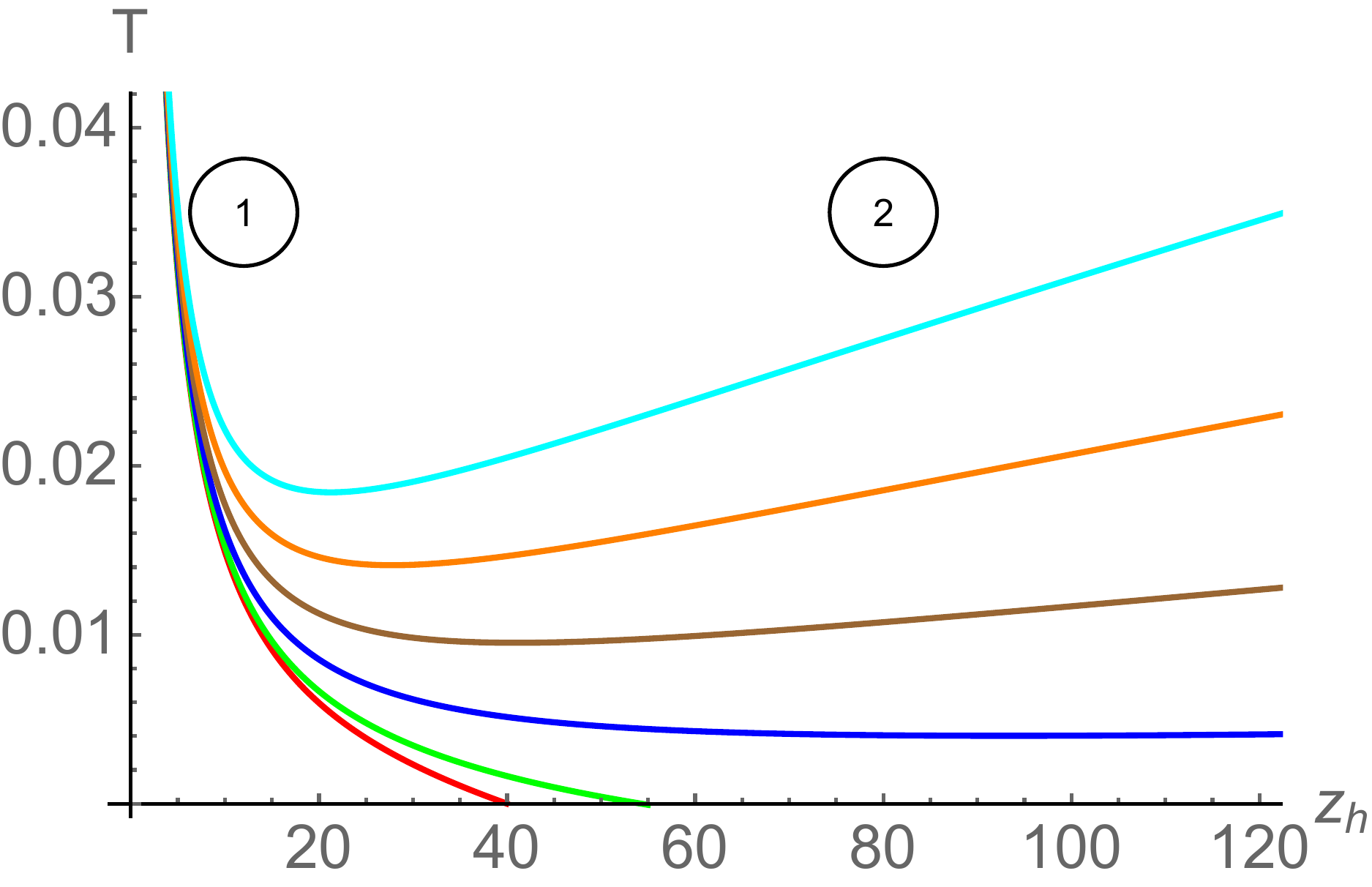}
\caption{ \small Hawking temperature $T$ as a function of horizon radius $z_h$ for various values of $a$.  Here $q_e=0.05$ is used. Red, green, blue, brown, orange, and cyan curves correspond to $a=0$, $0.02$, $0.04$, $0.06$, $0.08$, and $1.0$, respectively.}
\label{zhvsTvsaqPt05f1}
\end{minipage}
\hspace{0.4cm}
\begin{minipage}[b]{0.5\linewidth}
\centering
\includegraphics[width=2.8in,height=2.3in]{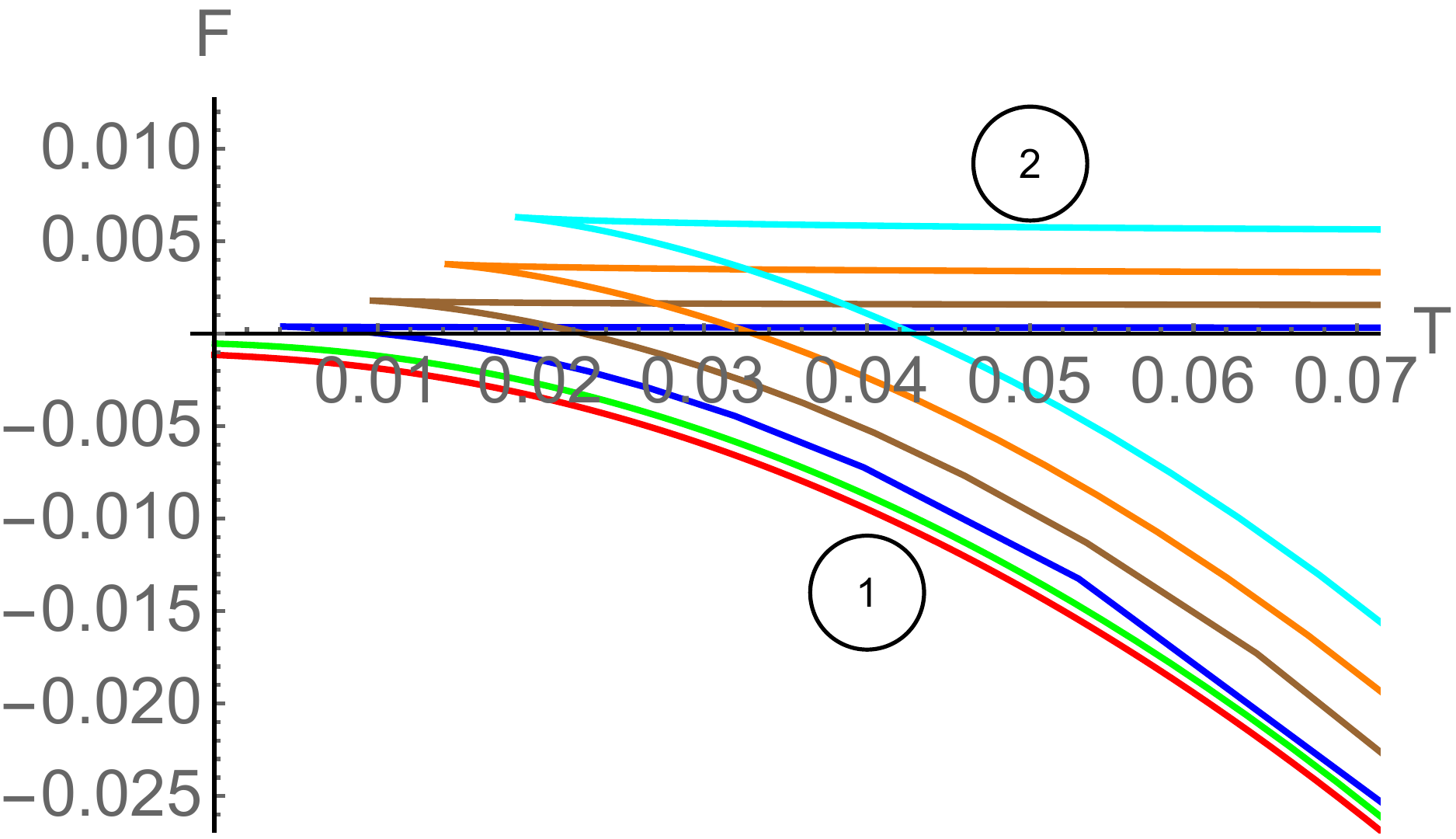}
\caption{\small Free energy $F$ as a function of Hawking temperature $T$ for various values of $a$. Here $q_e=0.05$ is used. Red, green, blue, brown, orange, and cyan curves correspond to $a=0$, $0.02$, $0.04$, $0.06$, $0.08$, and $0.10$, respectively.}
\label{TvsFvsaqPt05f1}
\end{minipage}
\end{figure}

The thermodynamic structure of the hairy black hole becomes even more interesting for finite charges. In particular, depending upon the relative magnitude of $q_e$ and $a$, not only the black hole can become extremal but also there can exist one or two black hole branches. This is shown in Fig.~\ref{zhvsTvsaqPt05f1}. The results here are shown for a particular value of $q_e = 0.05$; however, similar results occur for its other values as well. For $q_e \neq 0$ and small $a$, there exists only one stable black hole branch which becomes extremal at some horizon radius $z_h^{ext}$ (green line). This result is completely analogous to the charged BTZ black hole. For the charged BTZ black hole case, the extremal horizon radius occurs at $z_h^{ext}=2/q_e$, whereas for the hairy black hole case, the magnitude of this $z_h^{ext}$ increases with $a$. These results further imply that, irrespective of the temperature, at least one black hole branch always exists and remains stable for the charged case when $a$ is relatively small. We can also mathematically recognize the reason for the occurrence of the extremal black hole. Notice that only the third term in Eq.~(\ref{tempf1}) gives a negative contribution to the temperature. If this negative contribution is strong enough then it can make the black hole extremal. This is exactly what happens when $a$ is small.

However, for large $a$ values, keeping $q_e$ fixed, the temperature starts increasing with $z_h$ and the black hole never become extremal (brown line), i.e., the negative third term in Eq.~(\ref{tempf1}) is now always less than the sum of other two positive terms for relatively large values of $a$. Hence, just like in the case of $\{q_e=0$, $a>0\}$,  we again have two black hole branches. These black hole branches similarly exist only above $T_{min}$. Accordingly, there appears a black hole/thermal-AdS Hawking/Page phase transition at some transition temperature $T_{HP}$. The free energy behavior of the charged case is shown in Fig.~\ref{TvsFvsaqPt05f1}. Here again, the free energy changes sign at $T_{HP}$, thereby clearly showing the thermal-AdS/black hole phase transition at this temperature. This interesting result should be contrasted with the charged BTZ case, where no such phase transition appears. Moreover, this transition temperature is now a $q_e$ and $a$ dependent quantity. In particular, the transition temperature increases with $a$ whereas it decreases with $q_e$. The overall dependence of $T_{HP}$ on these parameters is shown in Fig.~\ref{avsTHPvsqf1}. Note that although $T_{HP}$ increases with $a$ for all $q_e$, however, unlike the $q_e=0$ case, the slope of $a$ vs $T_{HP}$ line is not constant for $q_e \neq 0$.

\begin{figure}[h!]
\centering
\includegraphics[width=2.8in,height=2.3in]{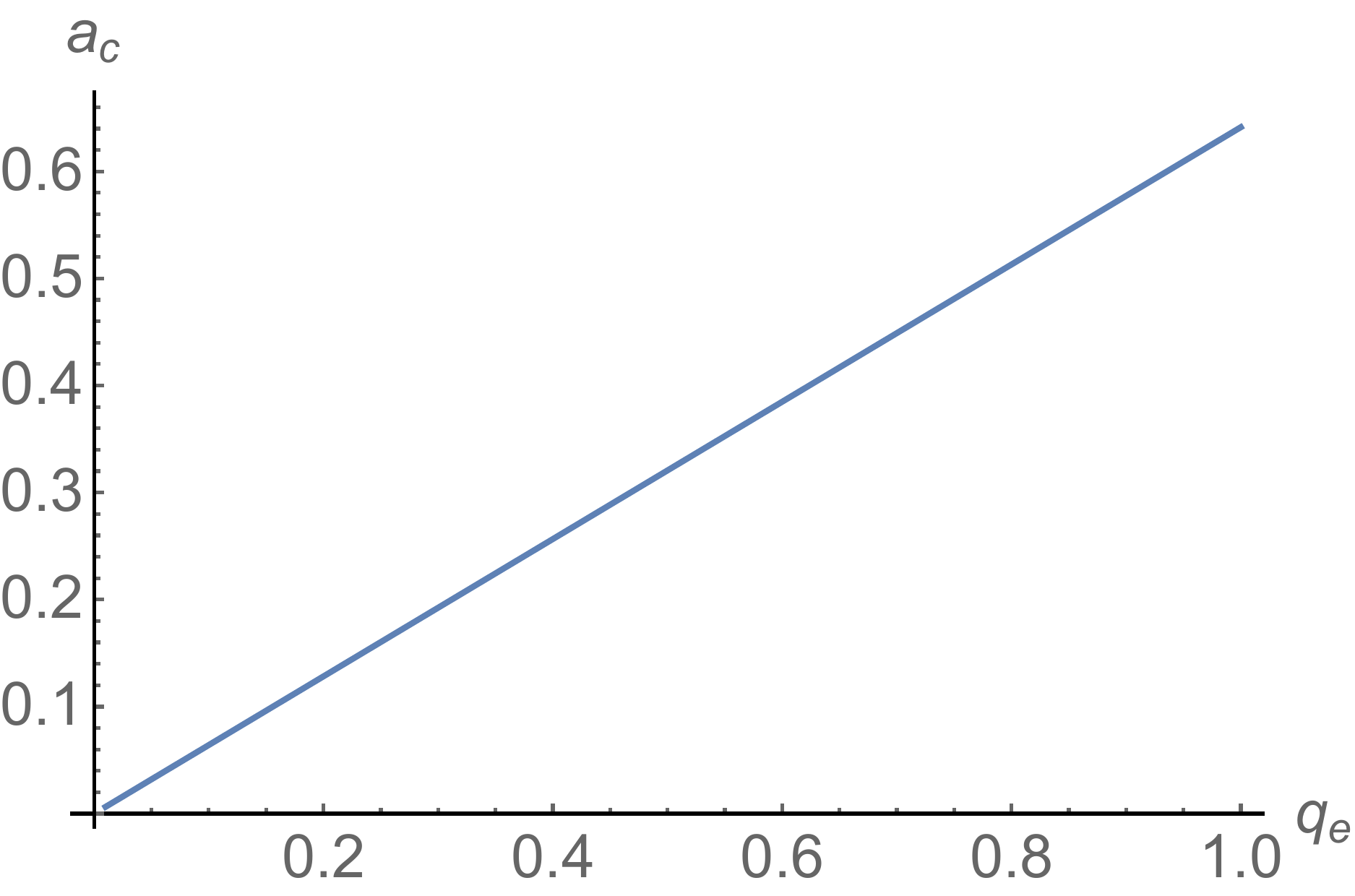}
\caption{ \small The variation of $a_c$ as a function of $q_e$.}
\label{qvsacf1}
\end{figure}

Our whole analysis above suggests that, for a fixed charge $q_e$, there exists a critical value for the hairy parameter $a=a_{c}$  above which the charged hairy black hole exhibits the Hawking/Page phase transition whereas below $a_c$ no such phase transition occurs. We further investigate how this $a_c$ varies with $q_e$. The result is presented in Fig.~\ref{qvsacf1}. We find that the critical magnitude of the hairy parameter increases linearly with $q_e$, suggesting more and more backreaction of the scalar hair is required for the larger charge black hole to undergo a phase transition. In fact, we can also obtain a relation between $a_c$ and $q_e$ by fitting the data of Fig.~\ref{qvsacf1}. Doing this, we get $a_c=0.6413 q_e$.

We again like to emphasize that the presence of Hawking/Page phase transition in the hairy case should be contrasted from the nonhairy charged BTZ case whereas no such phase transition occurs. \footnote{The Hawking/Page phase transition has also been found in a particular three-dimensional hairy gravity model in \cite{Zou:2014gla}. However, this phase transition appears only in the presence of angular momentum and for the non-rotating hairy case, like the one considered here, no such phase transition was observed.}

It is also important to discuss the local stability of these hairy black holes. The local stability is determined by the response of the equilibrium system under a small fluctuation in thermodynamical variables. In the canonical ensemble, it is quantified by the positivity of the specific heat at constant charge $C_{q_e}=T(\partial S_{BH}/\partial T)$. From the $z_h$ dependence of temperature in Figs.~\ref{zhvsTvsaq0f1} and \ref{zhvsTvsaqPt05f1} and from the fact that $S_{BH} \propto z_{h}^{-1}$, it is straightforward to observe that the slope of the $S_{BH}-T$ plane in the thermodynamically favored hairy black hole branch \textcircled{1}  is always positive. This in turn implies that $C_{q_e}>0$ in this branch, indicating the local stability of hairy black holes. Similarly, $C_{q_e}<0$ in the thermodynamically disfavored hairy black hole branch \textcircled{2}.

\section{HAIRY BLACK HOLES WITH $f(\phi)=e^{-\phi}$ COUPLING}
\begin{figure}[ht]
	\subfigure[]{
		\includegraphics[scale=0.4]{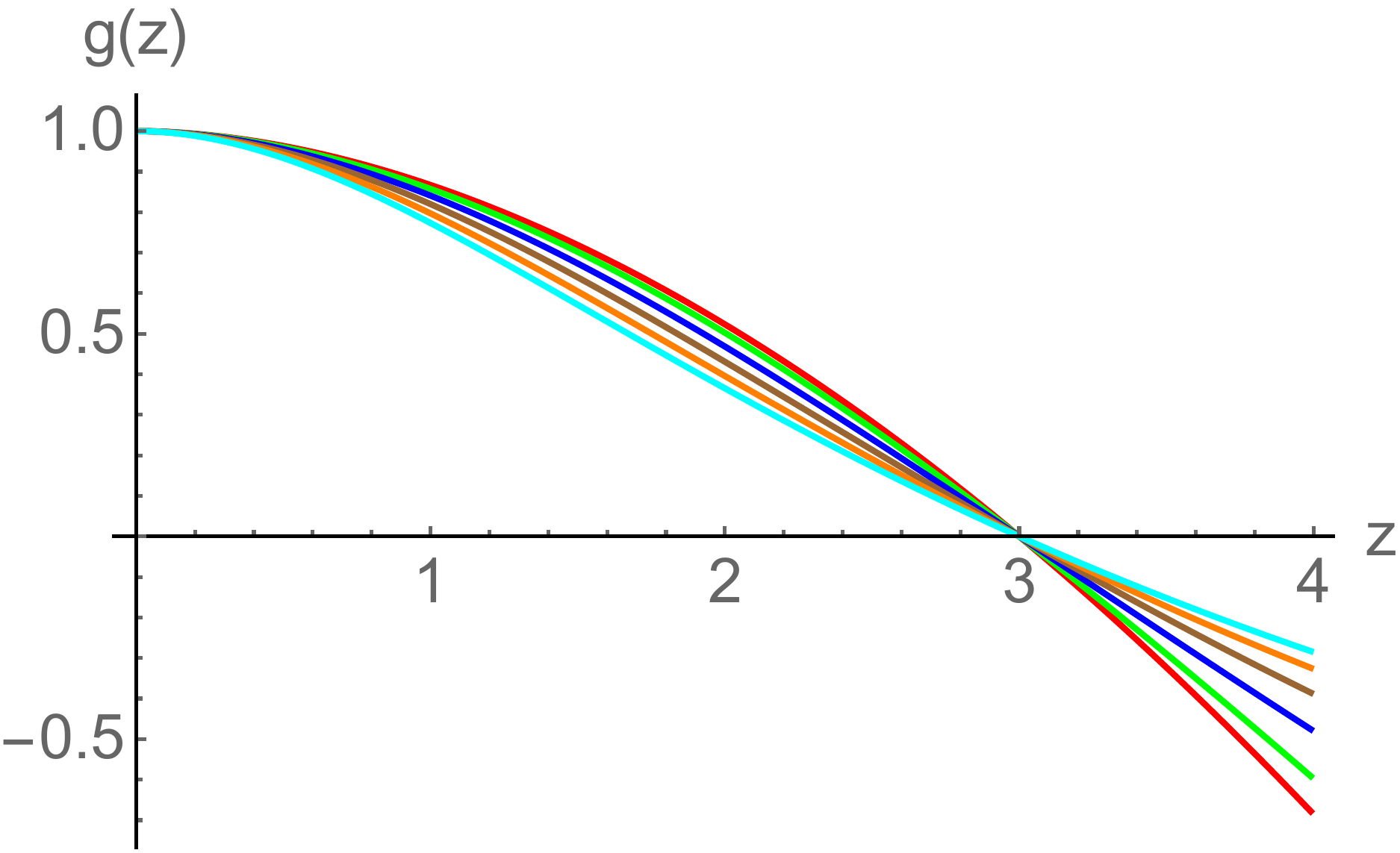}
	}
	\subfigure[]{
		\includegraphics[scale=0.4]{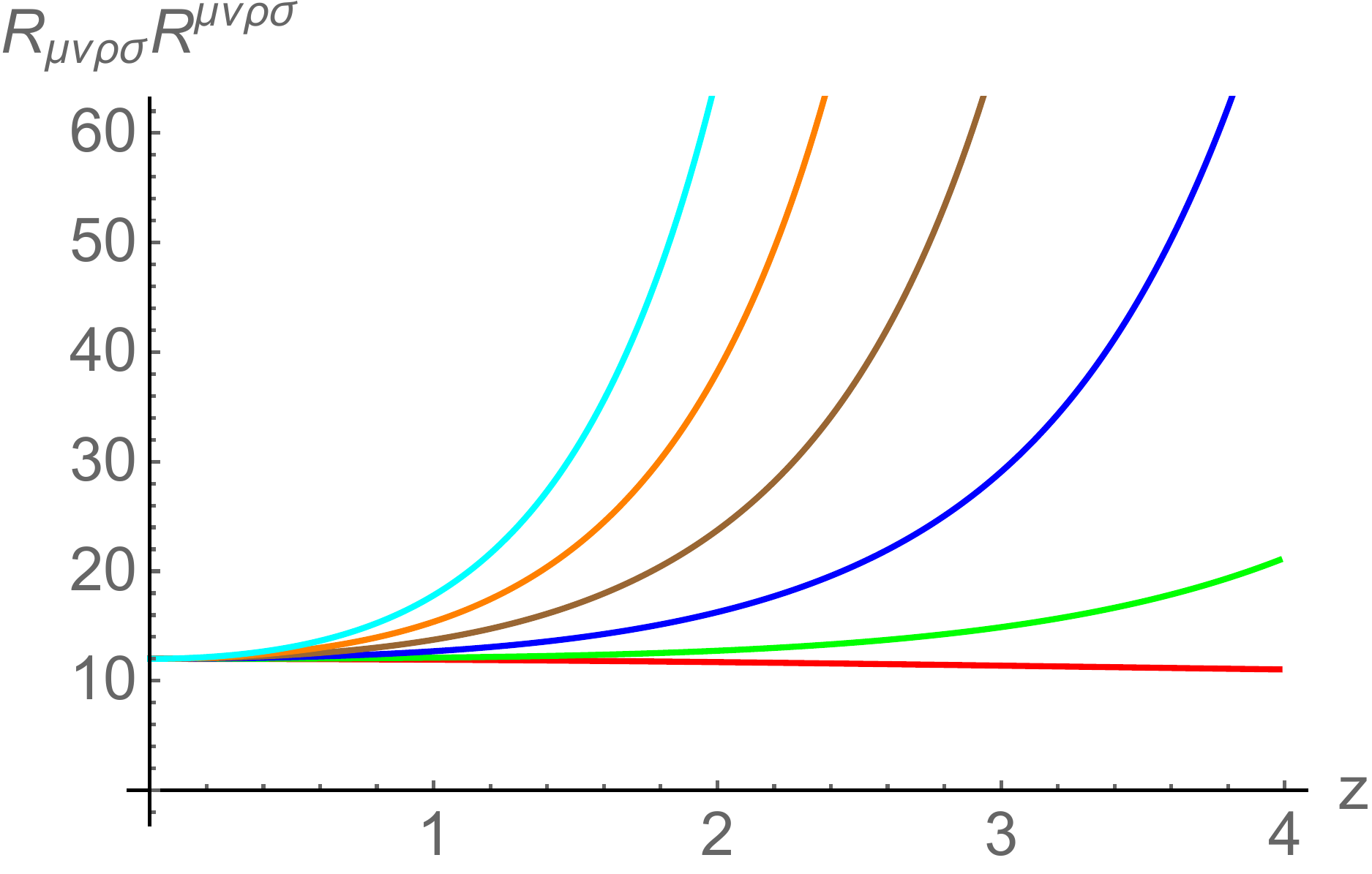}
	}
	\caption{\small The behavior of $g(z)$ and $R_{MNPQ}R^{MNPQ}$ for different values of hair parameter $a$. Here $z_h=3$ and $q_e=0.2$ are used. Red, green, blue, brown, orange, and cyan curves correspond to $a=0$, $0.1$, $0.2$, $0.3$, $0.4$, and $0.5$, respectively.}
	\label{zvsgzvsazh3qPt2f2}
\end{figure}

In this section, we move on to examine the geometrical and thermodynamical properties of the charged hairy black hole solution for the exponential coupling $f(\phi)=e^{-\phi}$. For $f(\phi)=e^{-\phi}$ and $A(z)=-\log{(1+a^2 z^2)}$, the solution for the scalar field will remain the same as in Eq.~(\ref{phibtsolf1}). This implies that the scalar field continues to be regular, finite, and well-behaved everywhere in the exterior horizon region for this exponential coupling function as well. The solution of the gauge field now reduces to
\begin{eqnarray}
& & B_t(z) = \frac{q_e}{2}  \left(\log \left(a^2 z^4+z^2\right)+4 \sinh ^{-1}(a z)\right)\,.
\label{Btsolf2}
\end{eqnarray}
Similarly, we get the following solution for $g(z)$
\begin{eqnarray}
& & g(z)=1 +\frac{\pi ^2 q_{e}^2}{12 a^2} + \frac{q_{e}^2 \log \left(1+a^2 z^2\right) \log \left(a^2 z^6+z^4\right)}{8 a^2}-\frac{\left(12 a^2+\pi ^2 q_{e}^2\right) \log \left(1+a^2 z^2\right)}{12 a^2 \log \left(1+a^2 z_h^2\right)} \nonumber\\
& & + \frac{q_{e}^2 \log \left(1+a^2 z^2\right) \sinh ^{-1}\left(a z_h\right) \left(\sinh ^{-1}\left(a
   z_h\right)-2 \log \left(e^{2 \sinh ^{-1}\left(a z_h\right)}+1\right)\right)}{a^2 \log \left(1+a^2
   z_h^2\right)} \nonumber\\
& &    -\frac{q_{e}^2 \log \left(1+a^2 z^2\right) \log \left(a^2 z_h^6+z_h^4\right)}{8 a^2} -\frac{q_{e}^2 \sinh ^{-1}(a z) \left(\sinh ^{-1}(a z)-2 \log \left(e^{2 \sinh ^{-1}(a
   z)}+1\right)\right)}{a^2}  \nonumber\\
& &   -\frac{q_{e}^2 \log \left(1+a^2 z^2\right) \text{Li}_2\left(-a^2 z_h^2\right)}{4 a^2 \log \left(1+a^2
   z_h^2\right)}-\frac{q_{e}^2 \log \left(1+a^2 z^2\right) \text{Li}_2\left(-e^{2 \sinh ^{-1}\left(a
   z_h\right)}\right)}{a^2 \log \left(1+a^2 z_h^2\right)} \nonumber\\
 & & +\frac{q_{e}^2 \text{Li}_2\left(-a^2 z^2\right)}{4
   a^2}+\frac{q_{e}^2 \text{Li}_2\left(-e^{2 \sinh ^{-1}(a z)}\right)}{a^2} \,.
\end{eqnarray}
These complicated-looking expressions of $B_t(z)$ and $g(z)$ again reduce to the standard charged BTZ black hole expressions in the limit $a\rightarrow 0$, hence once again highlighting the consistency of the obtained solution. In Fig.~\ref{zvsgzvsazh3qPt2f2}, the behavior of $g(z)$ and the Kretschmann scalar is shown. We see that spacetime exhibits a horizon at $z_h$ and is devoid of any additional singularity, thereby illustrating the smooth and well-behaved nature of the constructed hairy solution. The Kretschmann scalar again increases with $a$, showing that the spacetime becomes more curved as the backreaction of the hair increases. The same is true for the Ricci scalar. Similarly, the potential asymptotes to a constant value $V(z)|_{z\rightarrow 0} = 2\Lambda$ at the AdS boundary and is bounded from above.

Let us now discuss the thermodynamics of the hairy black hole. For $f(\phi)=e^{-\phi}$, the temperature of the black hole is given by
\begin{eqnarray}
& & T = \frac{a^2 z_h}{2 \pi  \log \left(1+a^2 z_h^2\right)} -\frac{q_{e}^2\sqrt{1+ a^2 z_h^2} \left(e^{2 \sinh ^{-1}\left(a z_h\right)}-1\right) \sinh ^{-1}\left(a
   z_h\right)}{2 \pi  a \left(e^{2 \sinh ^{-1}\left(a z_h\right)}+1\right)}  \nonumber \\
& & + \frac{q_{e}^2 z_h}{8 \pi} \left(\frac{\text{Li}_2\left(-a^2 z_h^2\right)+4 \text{Li}_2\left(-e^{2 \sinh ^{-1}\left(a
   z_h\right)}\right)}{\log \left(a^2 z_h^2+1\right)} +\frac{\pi^2}{3 \log \left(a^2 z_h^2+1\right)} - \frac{ \log \left(a^2
   z_h^2+1\right)}{2} \right) \nonumber \\
& & - \frac{q_{e}^2 z_h}{2 \pi} \left( \frac{\sinh ^{-1}\left(a z_h\right){}^2}{ \log \left(a^2
   z_h^2+1\right)} - \frac{2 \sinh ^{-1}\left(a z_h\right) \log \left(e^{2 \sinh ^{-1}\left(a
   z_h\right)}+1\right)}{\log \left(a^2 z_h^2+1\right)} \right) \,.
\label{tempfephi}
\end{eqnarray}
The above expression again reduces to the standard charged BTZ expression in the limit $a\rightarrow 0$. Notice that for $q_e=0$, the Einstein-Maxwell-scalar gravity action becomes identical for both $f(\phi)=1$ and $f(\phi)=e^{-\phi}$ couplings. This in turn implies that their thermodynamic structure also become identical. Accordingly, for the exponential coupling $f(\phi)=e^{-\phi}$ as well, there would again be a thermodynamically stable hairy black hole phase, with the possibility of phase transition to thermal-AdS as the temperature is lowered. In particular, the Hawking/page phase transition continues to exist, with a large stable hairy black hole phase dominating the phase structure at higher temperatures, whereas the thermal-AdS phase dominates the structure at lower temperatures.  For $q_e=0$, the phase diagram is essentially exactly similar to Fig.~\ref{avsTHPvsqf1} (red line).

\begin{figure}[h!]
\begin{minipage}[b]{0.5\linewidth}
\centering
\includegraphics[width=2.8in,height=2.3in]{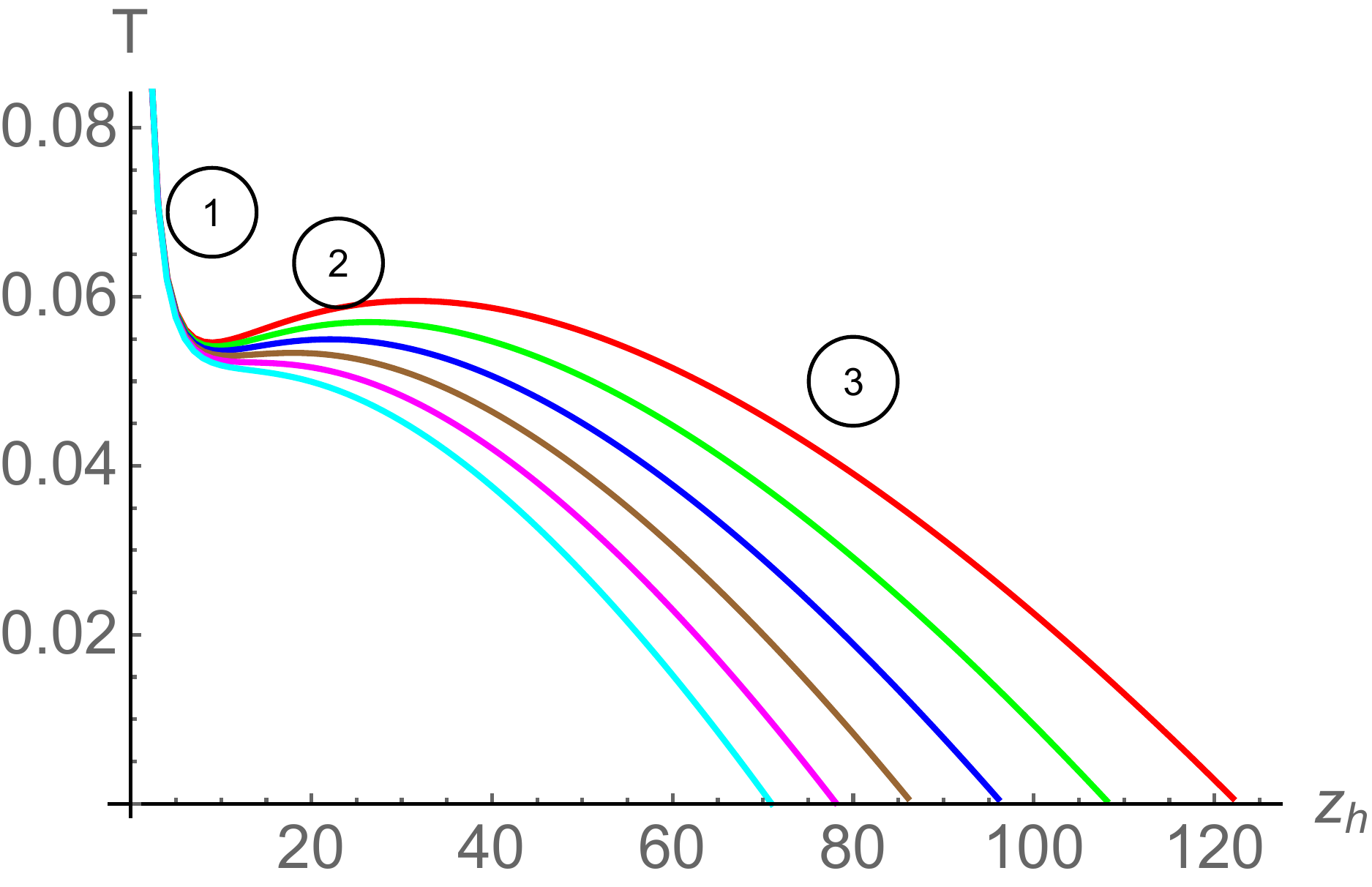}
\caption{ \small Hawking temperature $T$ as a function of horizon radius $z_h$ for various values of $q_e$.  Here $a=0.3$ is used. Red, green, blue, brown, orange, and cyan curves correspond to $q_e=0.058$, $0.06$, $0.062$, $0.064$, $0.066$, and $0.68$, respectively.}
\label{zhvsTvsqaPt3f2}
\end{minipage}
\hspace{0.4cm}
\begin{minipage}[b]{0.5\linewidth}
\centering
\includegraphics[width=2.8in,height=2.3in]{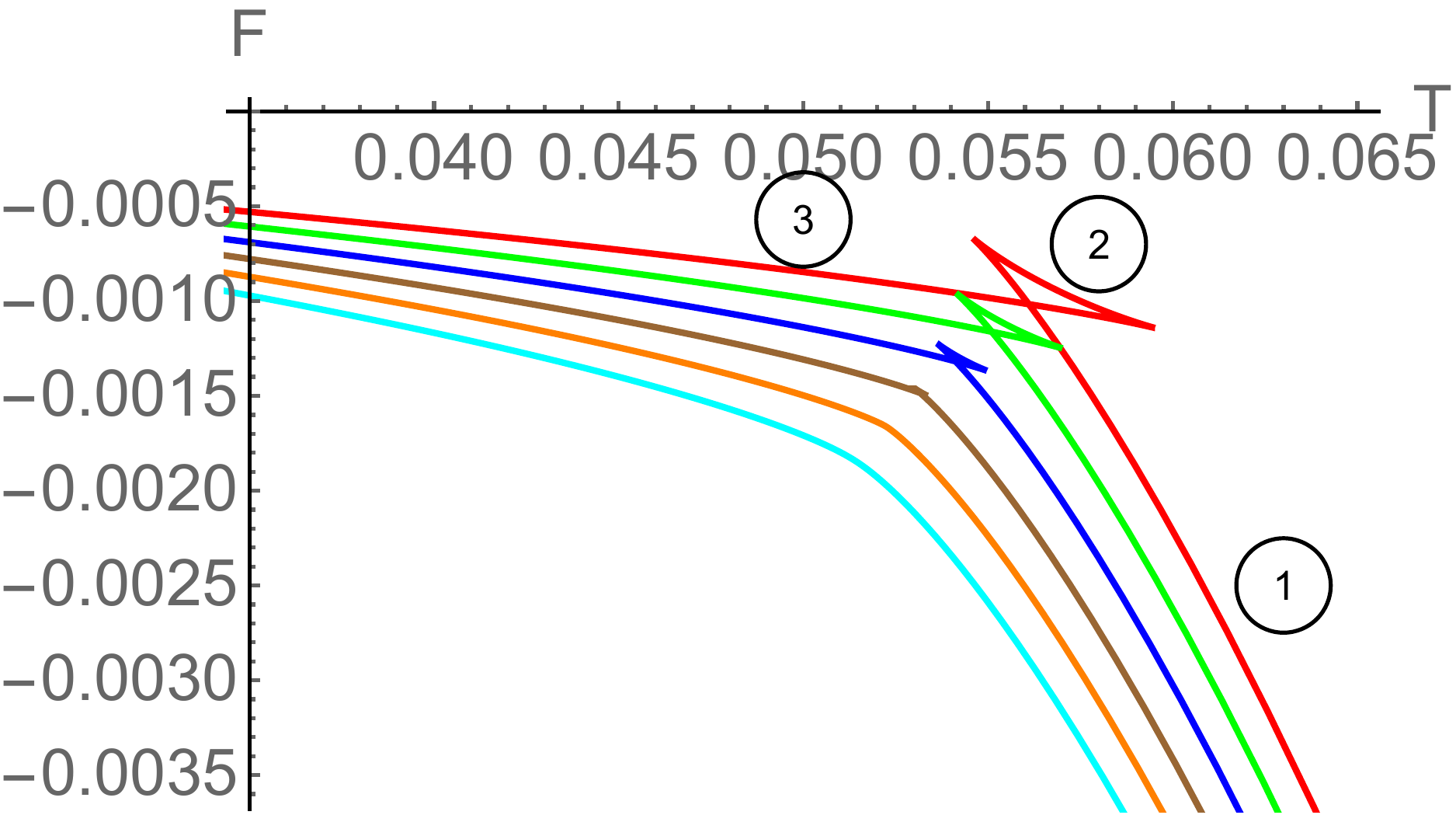}
\caption{\small Free energy $F$ as a function of Hawking temperature $T$ for various values of $q_e$. Here $a=0.3$ is used. Red, green, blue, brown, orange, and cyan curves correspond to $q_e=0.058$, $0.06$, $0.062$, $0.064$, $0.066$, and $0.068$, respectively.}
\label{TvsFvsqaPt3f2}
\end{minipage}
\end{figure}

The thermodynamic structure of the hairy black hole with $f(\phi)=e^{-\phi}$ coupling deviates from $f(\phi)=1$ coupling and becomes more interesting for finite charges. The results are shown in Figs.~\ref{zhvsTvsqaPt3f2} and \ref{TvsFvsqaPt3f2}. For a small but finite charge $0<q_e<q_{e}^{c}$, on top of a large stable black hole phase (marked by \textcircled{1}) and an unstable black hole phase (marked by \textcircled{2}), a new stable small black hole phase (marked by \textcircled{3}) now emerges with scalar hair at low temperatures. The temperature now has local minima and maxima and goes to zero at a finite radius $z_{h}^{ext}$. The small and large hairy black hole phases, for which the slope in the $(z_h-T)$ plane is negative, have positive specific heat $C_q$ and therefore are stable, whereas the intermediate phase, for which the slope is positive, has negative
specific heat and hence is unstable. Therefore, at least one stable black hole branch always exists at all temperatures. These results should be contrasted from the $f(\phi)=1$ case, where there were no such maxima in temperature and the stable small hairy black hole phase did not exist. Moreover, the magnitude of this $z_{h}^{ext}$ also depends nontrivially on $a$.

\begin{figure}[h!]
\begin{minipage}[b]{0.5\linewidth}
\centering
\includegraphics[width=2.8in,height=2.3in]{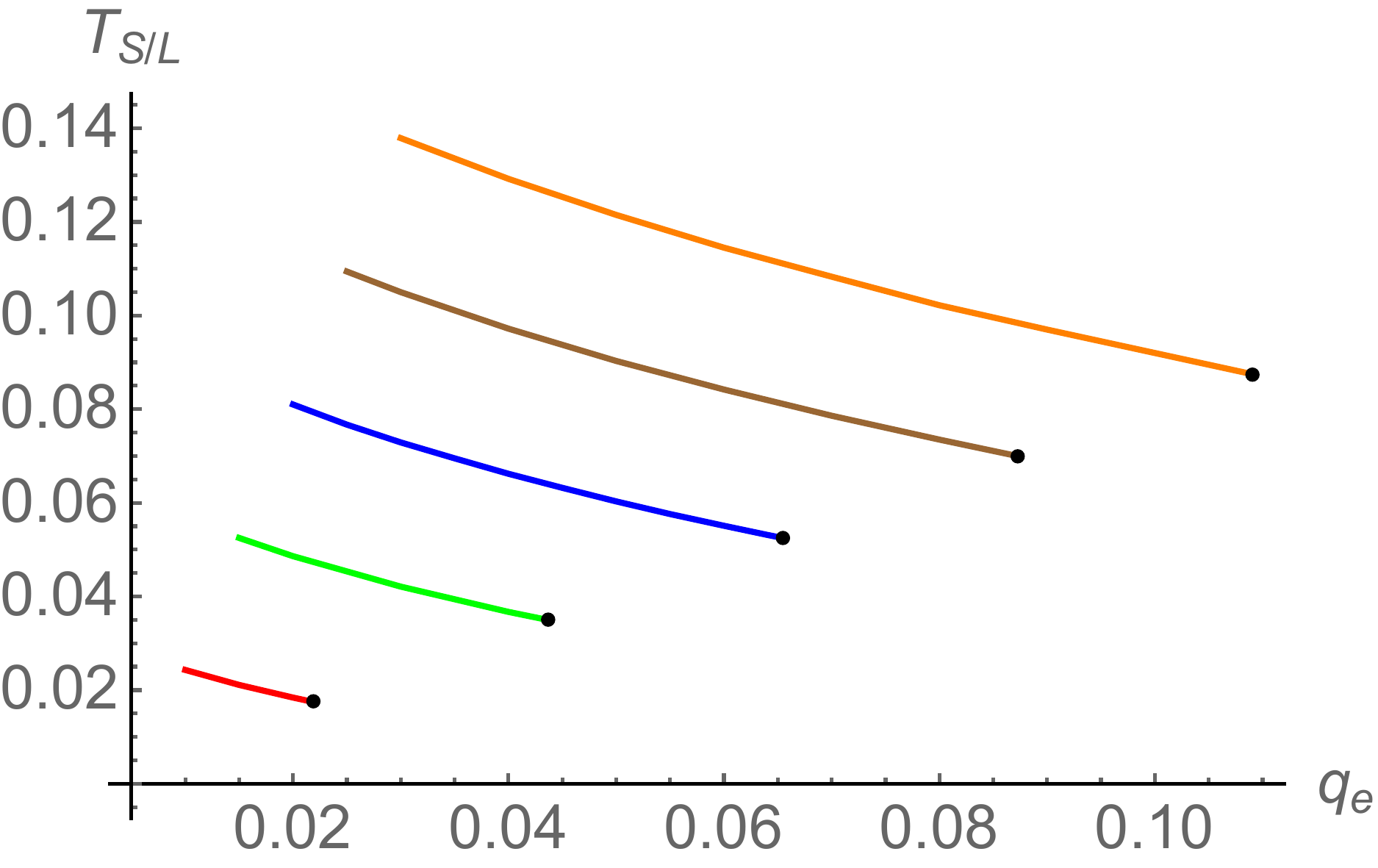}
\caption{ \small The transition temperature $T_{S/L}$ as a function of $q_e$ for different values of $a$. Red, green, blue, brown, and orange curves correspond to $a=0.1$, $0.2$, $0.3$, $0.4$, and $0.5$, respectively. Black dots indicate the second-order critical points $q_e^{c}$.}
\label{qevsTSLcritvsaf2}
\end{minipage}
\hspace{0.4cm}
\begin{minipage}[b]{0.5\linewidth}
\centering
\includegraphics[width=2.8in,height=2.3in]{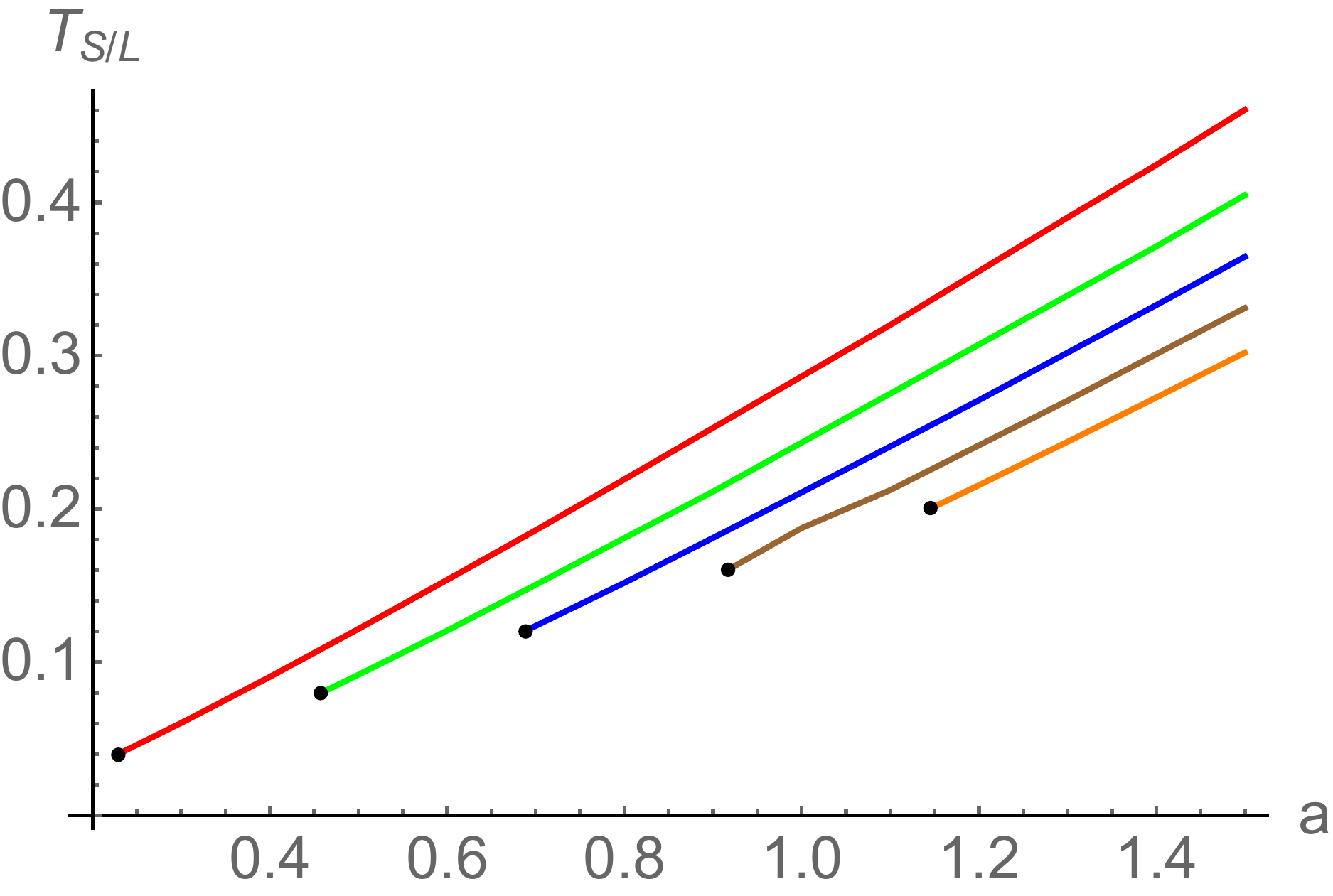}
\caption{\small The transition temperature $T_{S/L}$ as a function of $a$ for different values of $q_e$. Red, green, blue, brown, and orange curves correspond to $q_e=0.05$, $0.10$, $0.15$, $0.20$, and $0.25$, respectively. Black dots indicate the second-order critical points $a^{c}$.}
\label{avsTSLcritvsqef2}
\end{minipage}
\end{figure}

The multivaluedness of $z_h-T$ profile further suggests the possibility of phase transition between different black hole phases as the temperature is varied. The free energy behavior, shown in Fig.~\ref{TvsFvsqaPt3f2}, confirms this expectation. \footnote{Here the free energy is normalised with respect to the extremal case.} In particular, there appears a phase transition between the large black hole phase \textcircled{1}  and the small black hole phase \textcircled{3} as the temperature is varied. The free energy exhibits the standard swallowtail-like structure -- a characteristic feature of the first-order phase transition -- which exchanges dominance as the temperature is varied. In particular, the large black hole has a lower free energy at higher temperatures whereas the small black hole has lower free energy at lower temperatures. This indicates a phase transition from large to small hairy black holes (and vice versa) as the temperature is varied. Notice that the free energy of the unstable second phase \textcircled{2} is always higher than the stable first and third phases. Accordingly, it is always thermodynamically disfavored. The kink, where the free energy of large and small black holes becomes equal, defines a transition temperature $T_{S/L}$. We have further analyzed how this transition temperature depends on various black hole parameters. Our results are presented in Figs.~\ref{qevsTSLcritvsaf2} and \ref{avsTSLcritvsqef2}, showing a nontrivial dependence of $T_{S/L}$ on $q_e$ and $a$.

As the charge $q_e$ increases, the base of the swallowtail in free energy behavior starts decreasing and vanishes completely at a certain critical value of charge $q_e^{c}$. At $q_e^{c}$, the phase transition between large and small hairy black hole phases ceases to exist, and these two black hole phases combine together to form a single hairy black hole phase which remains stable at all temperatures. The critical charge $q_e^{c}$, therefore, describes a second-order critical point on which the first-order small/large black hole phase transition line terminates. The complete dependence of $T_{S/L}$ on $q_e$ and $a$ is shown in Figs.~\ref{qevsTSLcritvsaf2} and \ref{avsTSLcritvsqef2}.  Overall, the above thermodynamic behavior in the fixed charge ensemble resembles the famous van der Waals type phase transition, however now, in hairy black holes.

\begin{figure}[h!]
\begin{minipage}[b]{0.5\linewidth}
\centering
\includegraphics[width=2.8in,height=2.3in]{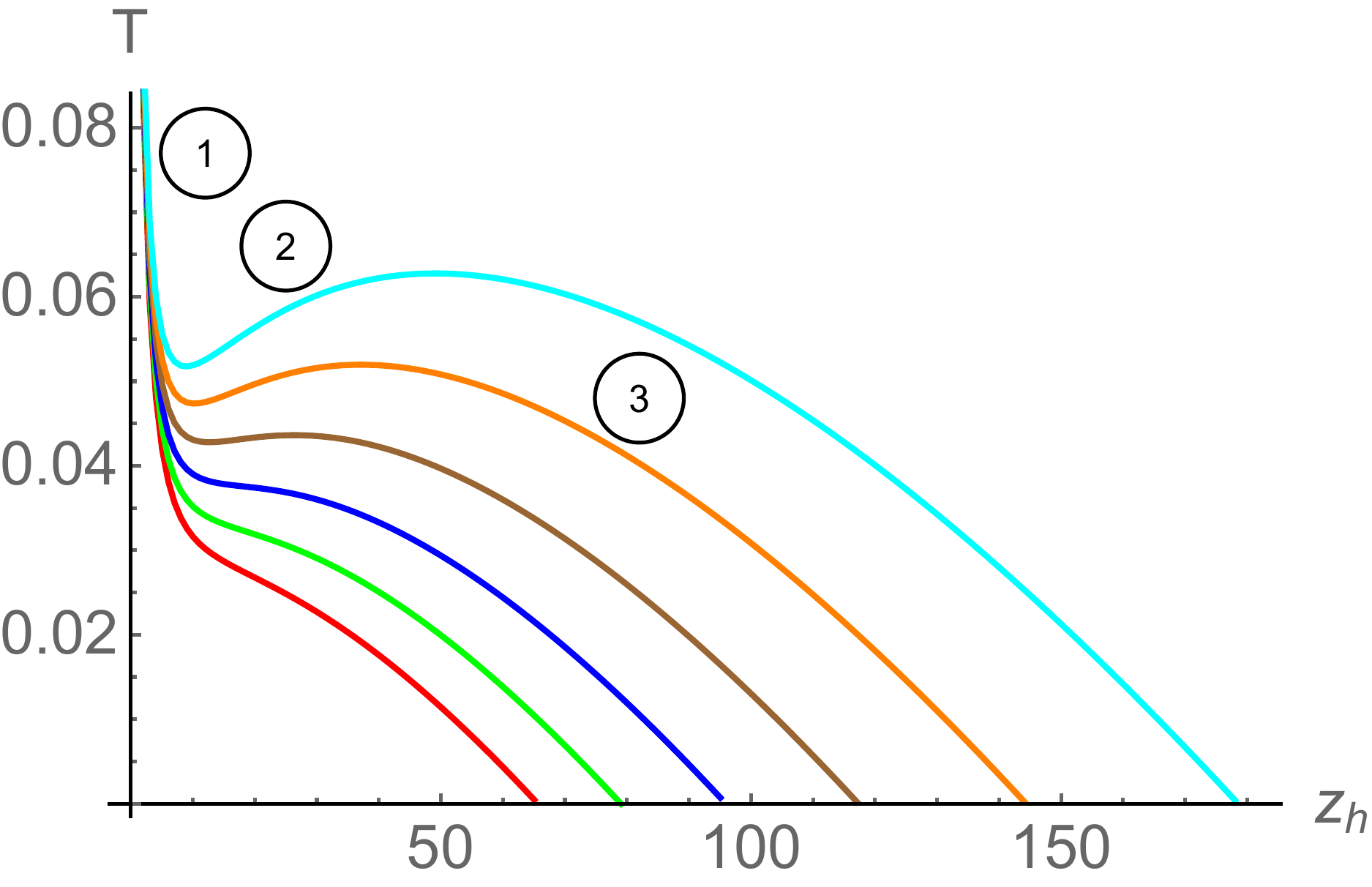}
\caption{ \small Hawking temperature $T$ as a function of horizon radius $z_h$ for various values of $a$.  Here $q_e=0.05$ is used. Red, green, blue, brown, orange, and cyan curves correspond to $a=0.18$, $0.20$, $0.22$, $0.24$, $0.26$, and $0.28$, respectively.}
\label{zhvsTvsaqPt05f2}
\end{minipage}
\hspace{0.4cm}
\begin{minipage}[b]{0.5\linewidth}
\centering
\includegraphics[width=2.8in,height=2.3in]{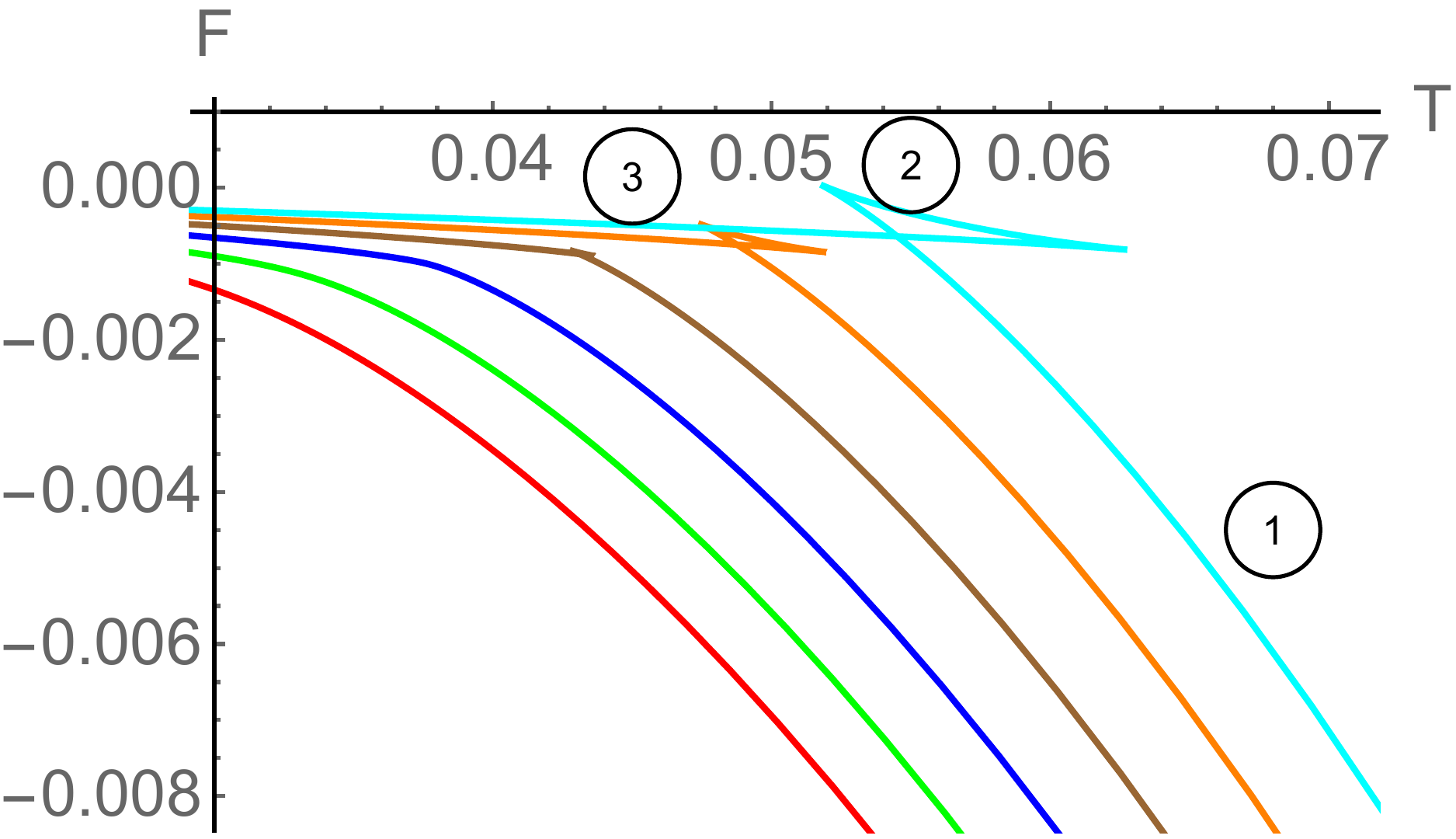}
\caption{\small Free energy $F$ as a function of Hawking temperature $T$ for various values of $a$. Here $q_e=0.05$ is used. Red, green, blue, brown, orange, and cyan curves correspond to $a=0.18$, $0.20$, $0.22$, $0.24$, $0.26$, and $0.28$, respectively.}
\label{TvsFvsaqPt05f2}
\end{minipage}
\end{figure}

\begin{figure}[h!]
\begin{minipage}[b]{0.5\linewidth}
\centering
\includegraphics[width=2.8in,height=2.3in]{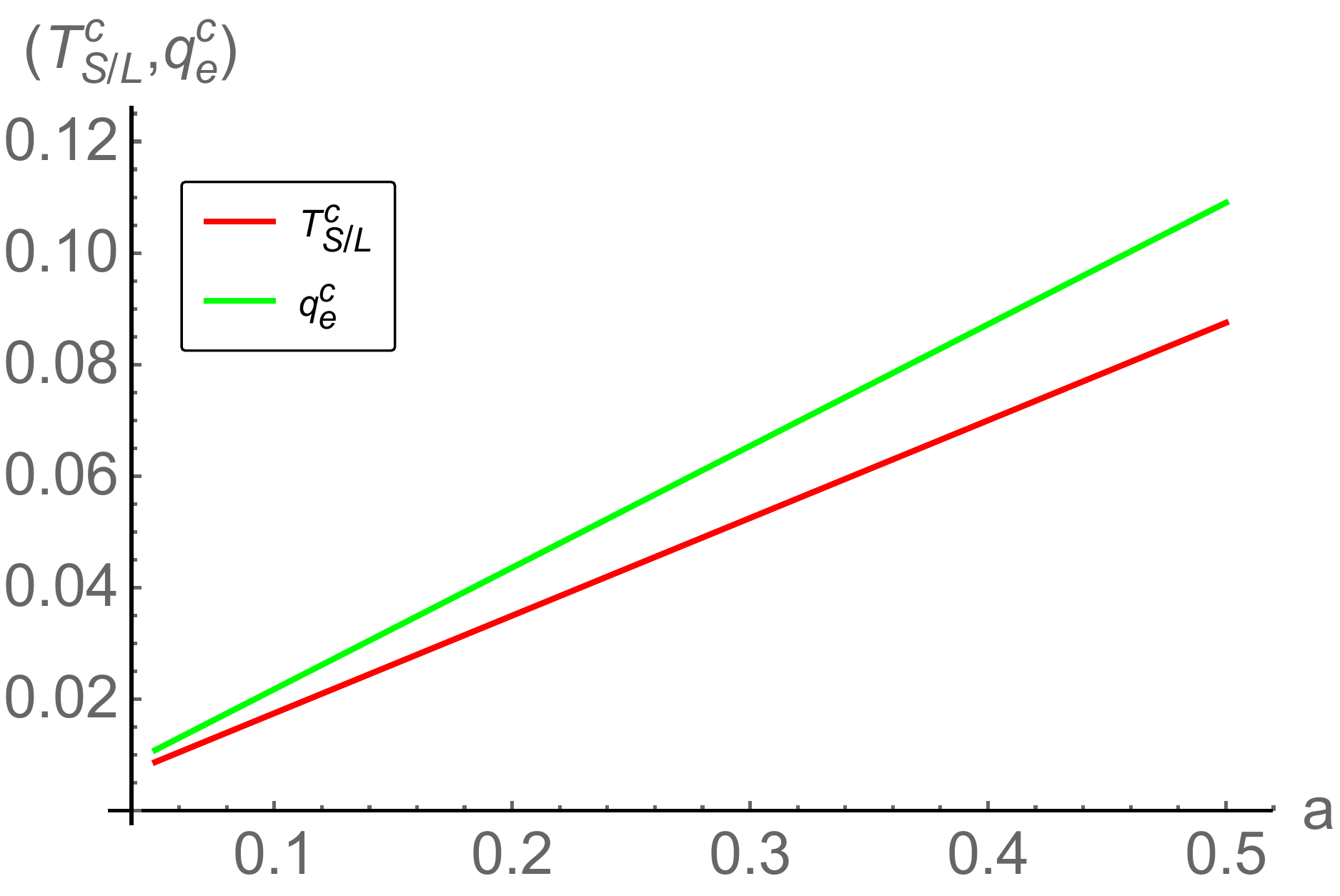}
\caption{ \small The Hawking temperature $T_{S/L}^{c}$ and charge $q_{e}^c$ at the second-order critical point as a function of $a$.}
\label{avsqecritvstcritf2}
\end{minipage}
\hspace{0.4cm}
\begin{minipage}[b]{0.5\linewidth}
\centering
\includegraphics[width=2.8in,height=2.3in]{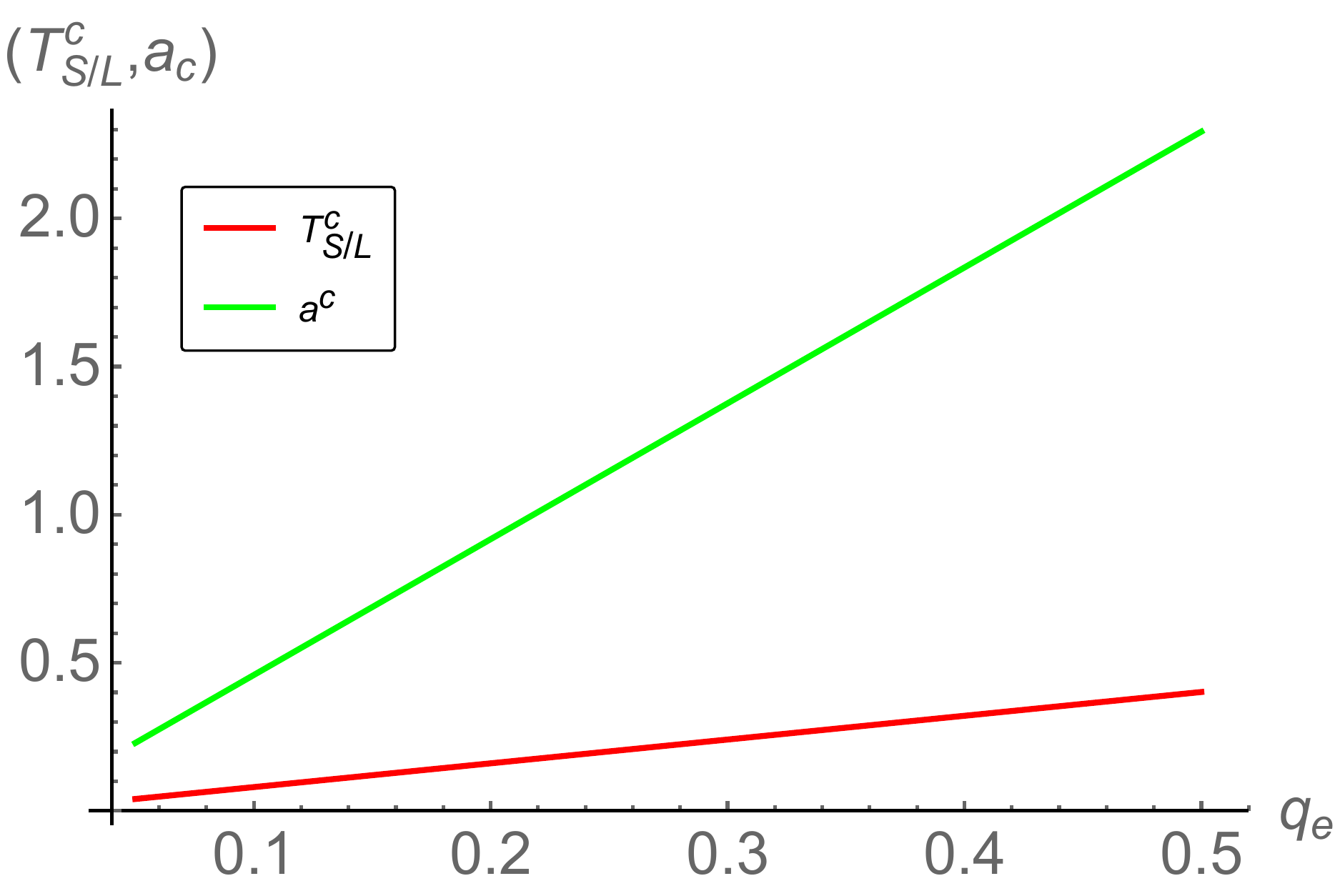}
\caption{\small The Hawking temperature $T_{S/L}^{c}$ and hair parameter $a_c$ at the second-order critical point as a function of $q_e$.}
\label{qevsacritvstcritf2}
\end{minipage}
\end{figure}

Similarly, the hairy black hole phase structure changes considerably by varying $a$ in the case of exponential coupling. For lower values of $a$, there is only one black hole phase. This is completely analogous to the charged BTZ case, albeit with a different extremal radius. However, for higher values of $a$, keeping $q_e$ fixed, now three black hole phases appear. This is shown in Figs.~\ref{zhvsTvsaqPt05f2} and \ref{TvsFvsaqPt05f2}. The stable first and third phases are always thermodynamically favored over the unstable second phase, and there is again a phase transition between the first and third phases at $T_{S/L}$. Therefore, the small/large black hole phase transition appears for higher values of $a$ whereas it ceases to exist at smaller values of $a$. This behavior is different from the varying charge scenario discussed above, where higher values of charge instead lead to the dissolution of small/large phase transition. These results further suggest that, like $q_e^{c}$, there also exists a critical value of the hair parameter ($a^c$) at which the first-order small/large black hole phase transition line stops. As usual, the magnitude of $q_e^{c}$ and $a^c$ can be obtained by analyzing the inflection point of temperature, i.e.,

\begin{eqnarray}
& & \frac{\partial T}{\partial z_h} = 0 = \frac{\partial T^2}{\partial z_{h}^2}  \,,
\end{eqnarray}
with $z_h=z_{h}^c$ and $q_e=q_e^{c}$ or $a=a^c$. The complete phase diagram showing the dependence of $T_{S/L}^{c}$ (the hawking temperature at the second-order critical point) on $q_e$ and $a$, as well as the critical points $q_e^{c}$ and $a^c$ are shown in Figs.~\ref{avsqecritvstcritf2} and \ref{qevsacritvstcritf2}.

At this point, it is important to emphasize that the usual charged BTZ black hole does not exhibit the small/large black hole phase transition. Here we find that, depending upon the relative magnitude of charge and hair parameter, the small/large black hole phase transition can take place in the canonical ensemble in the presence of scalar hair. To the best of our knowledge, such van der Waals type phase transition has not been observed in three dimensions in the literature before. As we show in the next section such small/large black hole phase transition persists for other exponential couplings as well, making such transitions a robust phenomenon in our gravity model.

\section{HAIRY BLACK HOLES WITH $f(\phi)=e^{-\phi^2/2}$ COUPLING}
\begin{figure}[ht]
	\subfigure[]{
		\includegraphics[scale=0.4]{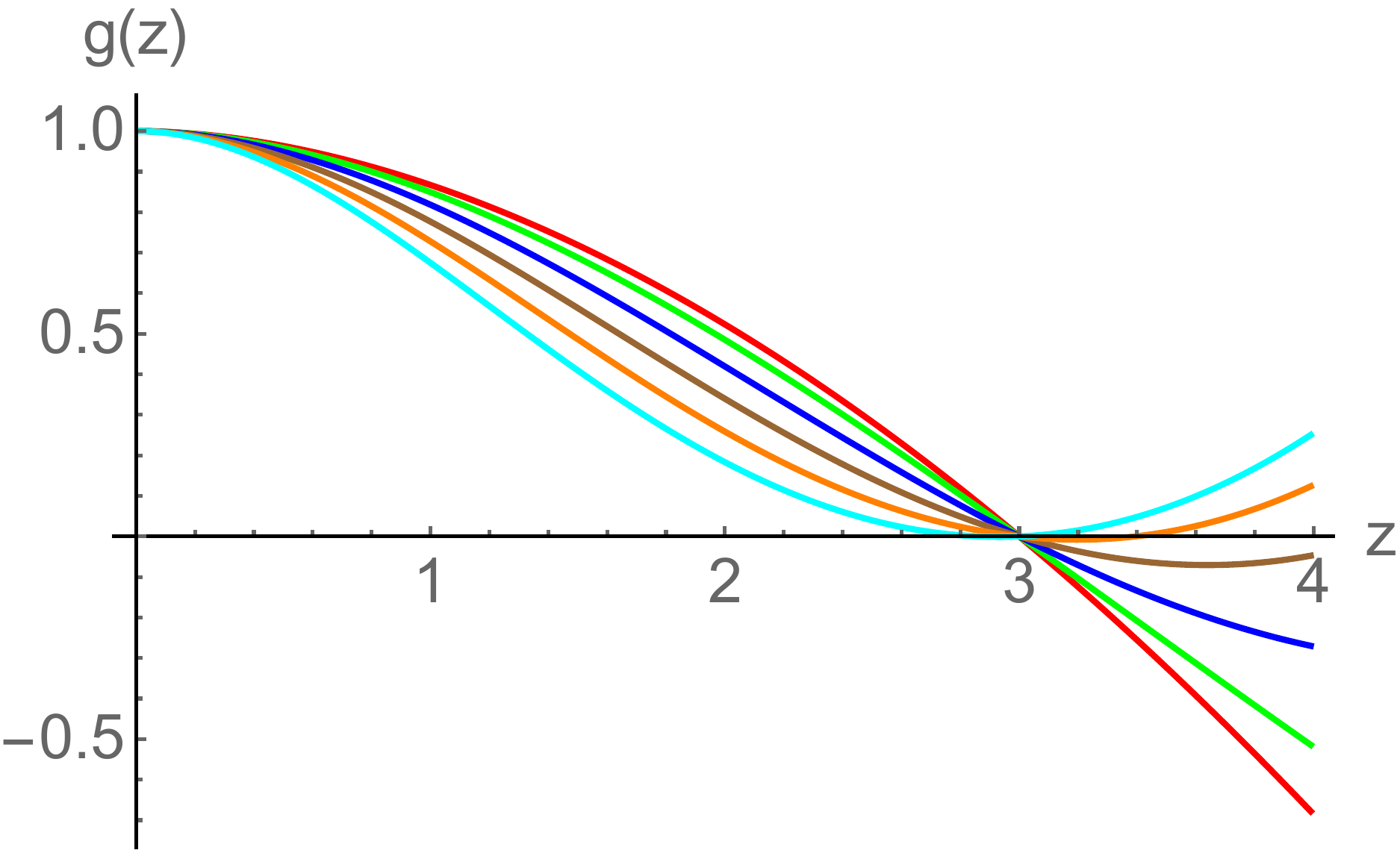}
	}
	\subfigure[]{
		\includegraphics[scale=0.4]{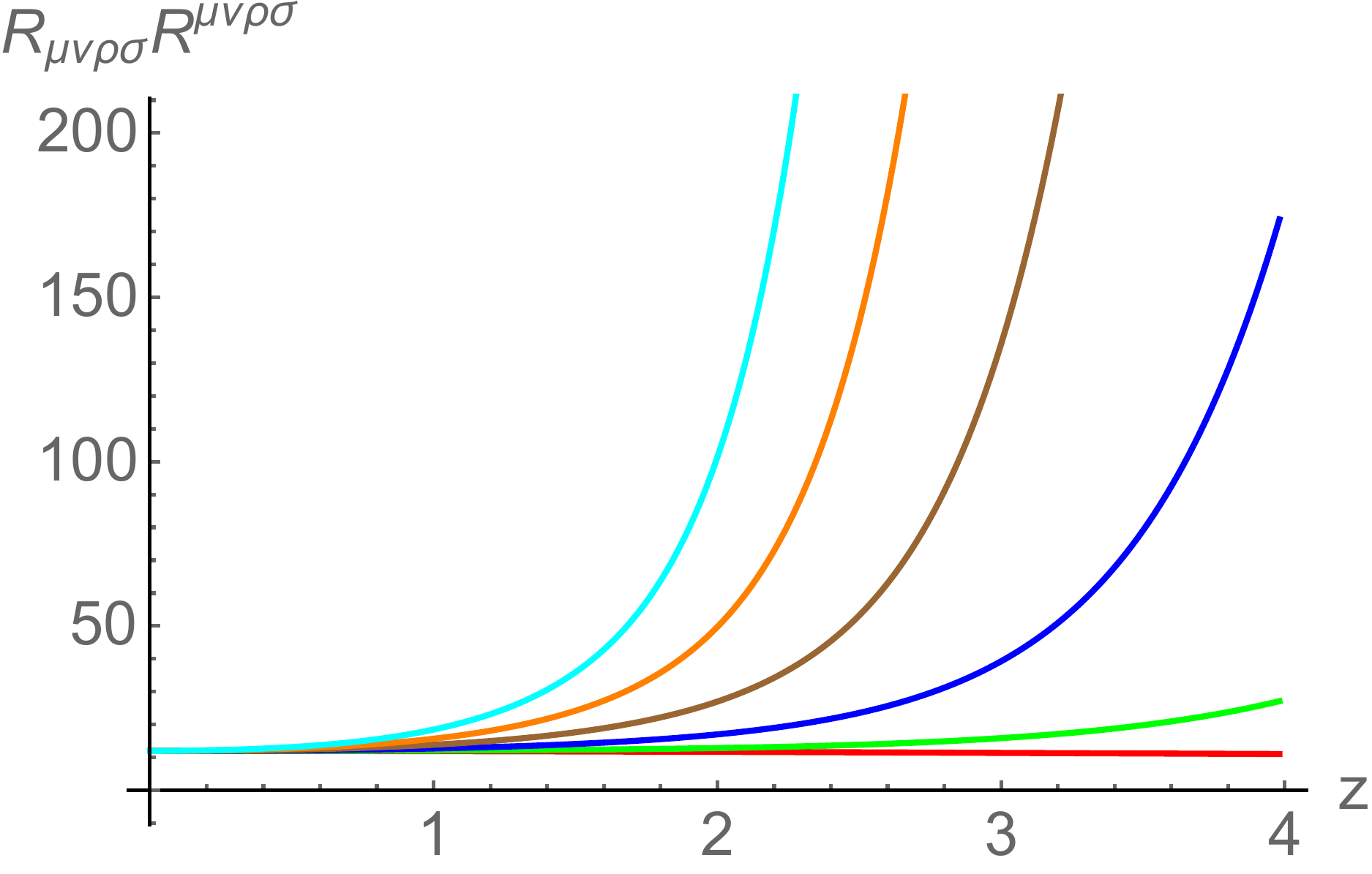}
	}
	\caption{\small The behavior of $g(z)$ and $R_{MNPQ}R^{MNPQ}$ for different values of hair parameter $a$. Here $z_h=3$ and $q_e=0.2$ are used. Red, green, blue, brown, orange, and cyan curves correspond to $a=0$, $0.1$, $0.2$, $0.3$, $0.4$, and $0.5$, respectively.}
	\label{zvsgzvsazh3qPt2f3}
\end{figure}

Next, we examine the charged hairy black hole properties for another exponential coupling $f(\phi)=e^{-\phi^2/2}$. Such a coupling has been used to construct interesting scalarized black holes in four dimensions \cite{Herdeiro:2018wub}, and therefore, it is also instructive to investigate this coupling in three dimensions. With $f(\phi)=e^{-\phi^2/2}$, most of our results for the hairy solution remain the same as in the previous case of $f(\phi)=e^{-\phi}$. We will, therefore, be brief here. The solution of the gauge field reduces to
\begin{eqnarray}
& & B_t(z) = q \left(4 a z \left(\sqrt{1+a^2 z^2}+a z\right)-\frac{1}{2} \log \left(\frac{1+a^2z^2}{z^2}\right)\right) \,.
\label{Btsolf3}
\end{eqnarray}
Similarly, the solution for $g(z)$ is
\begin{eqnarray}
& & g(z)= 1-\frac{\log \left(1+a^2 z^2\right)}{\log \left(1+a^2 z_h^2\right)} + q_{e}^2 \left(\frac{2 z \sqrt{1+a^2 z^2}}{a}-\frac{2 \sinh ^{-1}(a z)}{a^2}+2 z^2\right) \nonumber\\
& & + \frac{q_{e}^2 \log \left(1+a^2 z^2\right)}{8 a^2} \left(\log \left(\frac{z^4}{\left(1+a^2 z^2\right)
   z_h^4}\right)-\frac{16 a^2 z_h^2}{\log \left(1+a^2 z_h^2\right)}\right)  \nonumber\\
& & -\frac{q_{e}^2 \log \left(1+a^2 z^2\right) \left(16 a z_h \sqrt{1+a^2 z_h^2}-\log ^2\left(1+a^2 z_h^2\right)-16
   \sinh ^{-1}\left(a z_h\right)\right)}{8 a^2 \log \left(1+a^2 z_h^2\right)} \nonumber\\
& & + \frac{q_{e}^2 \text{Li}_2\left(-a^2 z^2\right)}{4 a^2}  -\frac{q_{e}^2 \log \left(1+a^2 z^2\right) \text{Li}_2\left(-a^2
   z_h^2\right)}{4 a^2 \log \left(1+a^2 z_h^2\right)}  \,.
\end{eqnarray}
The profile of $g(z)$ and Kretschmann scalar is shown in Fig.~\ref{zvsgzvsazh3qPt2f3}. The spacetime is again regular and well-behaved everywhere outside the horizon, with curvature singularity appearing only inside the horizon. The potential asymptotes to a constant value $V(z)|_{z\rightarrow 0} = 2\Lambda$ at the AdS boundary and is bounded from above. The scalar field is similarly finite and regular, establishing the well-behaved geometric nature of the hairy black hole.

Now, the temperature of the black hole is
\begin{eqnarray}
& &T = \frac{q_{e}^2 z_h}{8 \pi} \left( \frac{\left(\log \left(a^2 z_h^2+1\right)-16 \left(a z_h \left(\sqrt{a^2 z_h^2+1}+a
   z_h\right)+1\right)\right)}{2} +\frac{\text{Li}_2\left(-a^2 z_h^2\right)}{\log \left(a^2 z_h^2+1\right)} \right) \nonumber\\
& &+ \frac{8 a z_h \left(2 q^2 z_h \sqrt{a^2 z_h^2+1}+2 a q_{e}^2 z_h^2+a\right)-16 q^2 z_h \sinh ^{-1}\left(a
   z_h\right)}{16 \pi  \log \left(a^2 z_h^2+1\right)} \,.
\end{eqnarray}

\begin{figure}[h!]
\begin{minipage}[b]{0.5\linewidth}
\centering
\includegraphics[width=2.8in,height=2.3in]{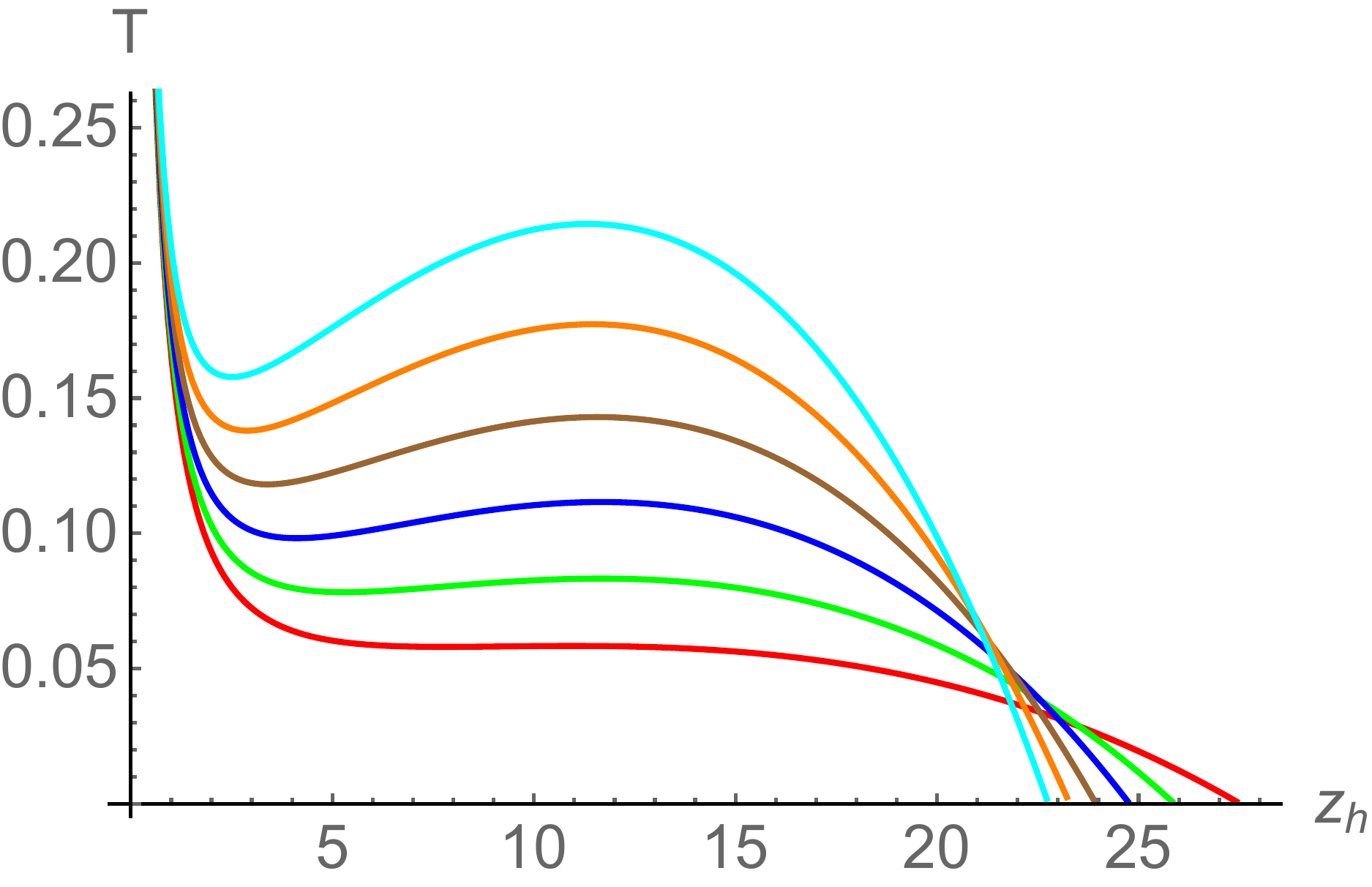}
\caption{ \small Hawking temperature $T$ as a function of horizon radius $z_h$ for various values of $a$.  Here $q_e=0.01$ is used. Red, green, blue, brown, orange, and cyan curves correspond to $a=0.3$, $0.4$, $0.5$, $0.6$, $0.7$, and $0.8$, respectively.}
\label{zhvsTvsaqPt01f3}
\end{minipage}
\hspace{0.4cm}
\begin{minipage}[b]{0.5\linewidth}
\centering
\includegraphics[width=2.8in,height=2.3in]{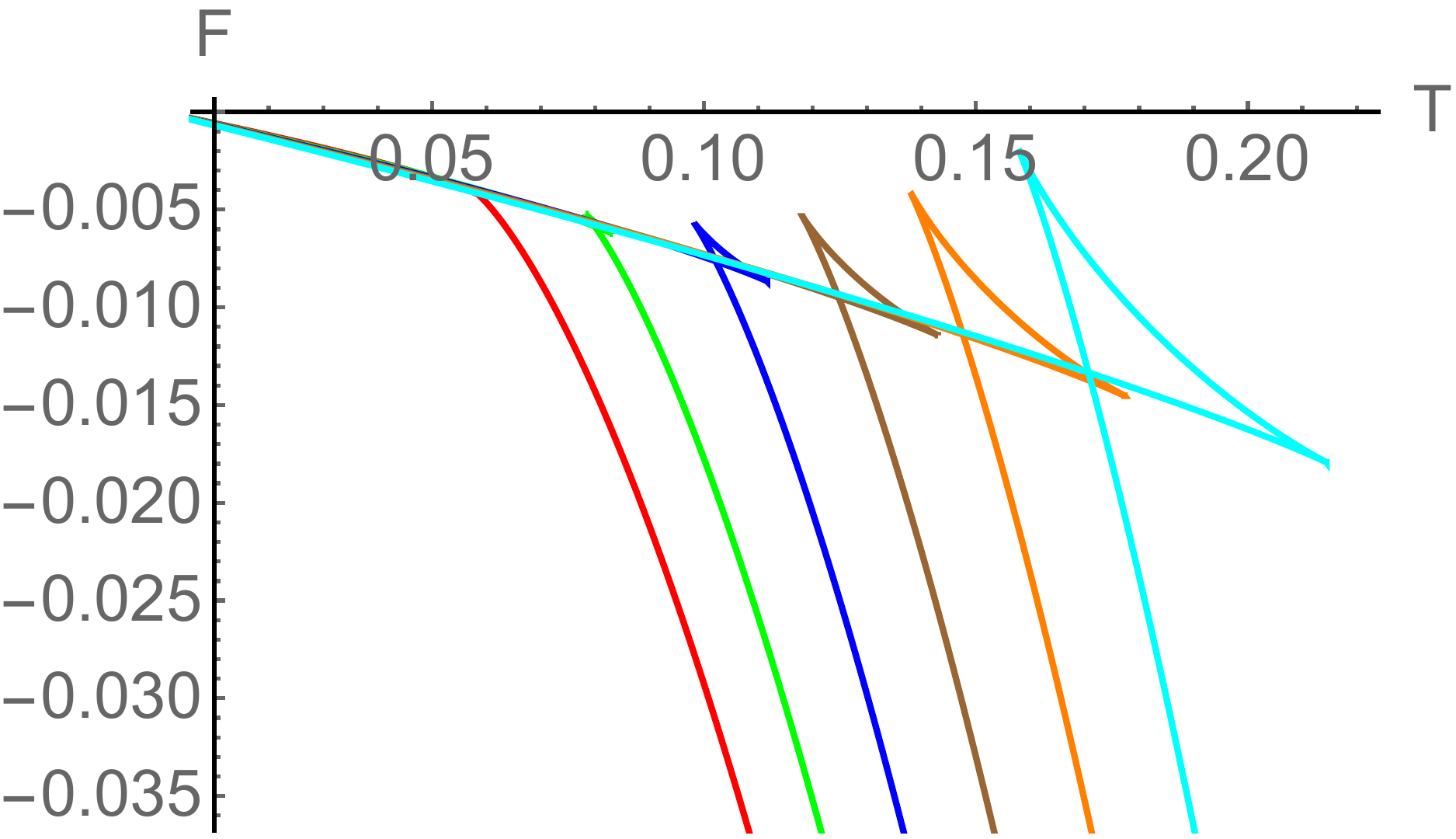}
\caption{\small Free energy $F$ as a function of Hawking temperature $T$ for various values of $a$. Here $q_e=0.01$ is used. Red, green, blue, brown, orange, and cyan curves correspond to $a=0.3$, $0.4$, $0.5$, $0.6$, $0.7$, and $0.8$, respectively.}
\label{TvsFvsaqPt01f3}
\end{minipage}
\end{figure}

\begin{figure}[h!]
\begin{minipage}[b]{0.5\linewidth}
\centering
\includegraphics[width=2.8in,height=2.3in]{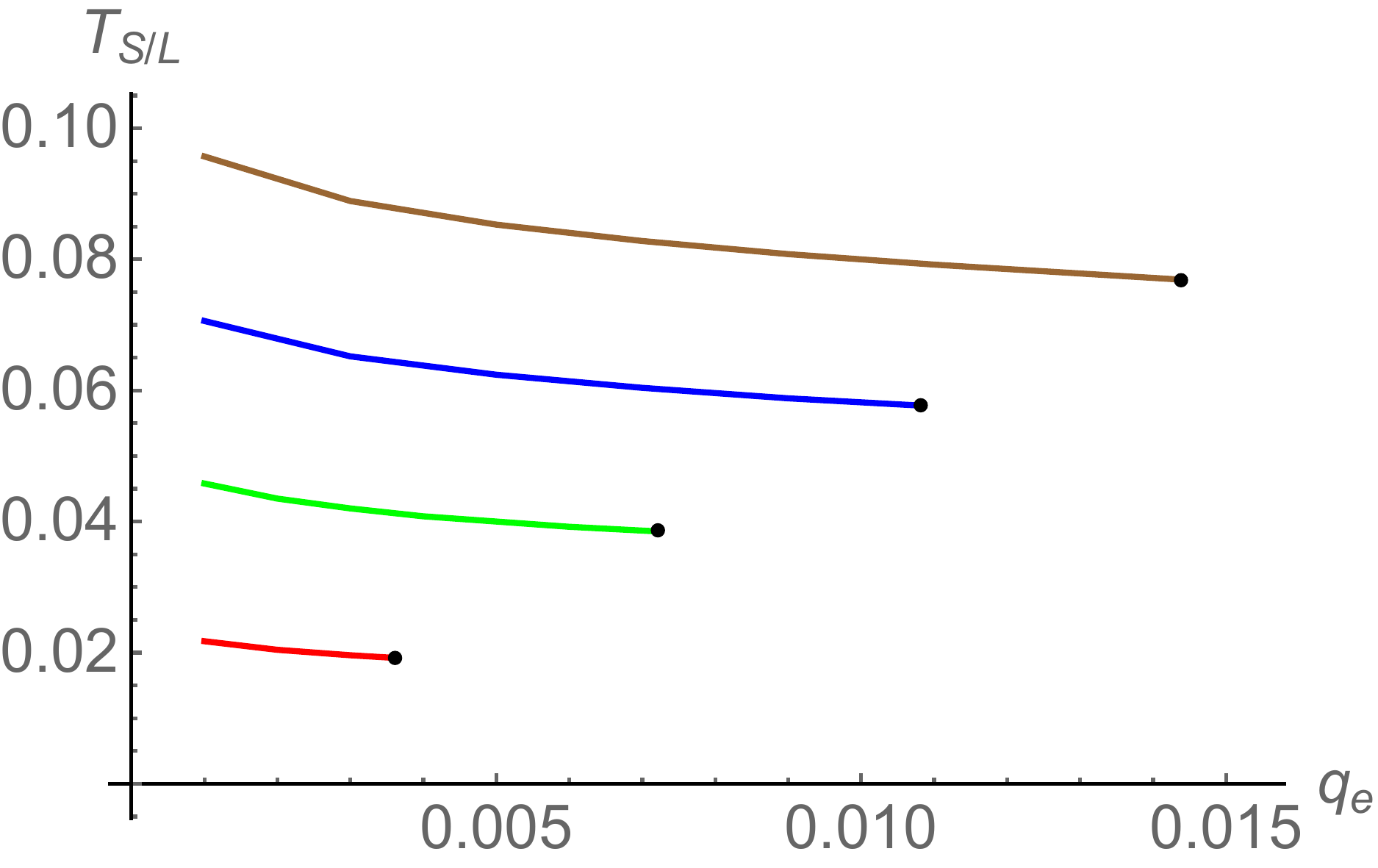}
\caption{ \small The transition temperature $T_{S/L}$ as a function of $q_e$ for different values of $a$. Red, green, blue, and brown curves correspond to $a=0.1$, $0.2$, $0.3$, and $0.4$, respectively. Black dots indicate the second-order critical points $q_e^{c}$.}
\label{qevsTSLcritvsaf3}
\end{minipage}
\hspace{0.4cm}
\begin{minipage}[b]{0.5\linewidth}
\centering
\includegraphics[width=2.8in,height=2.3in]{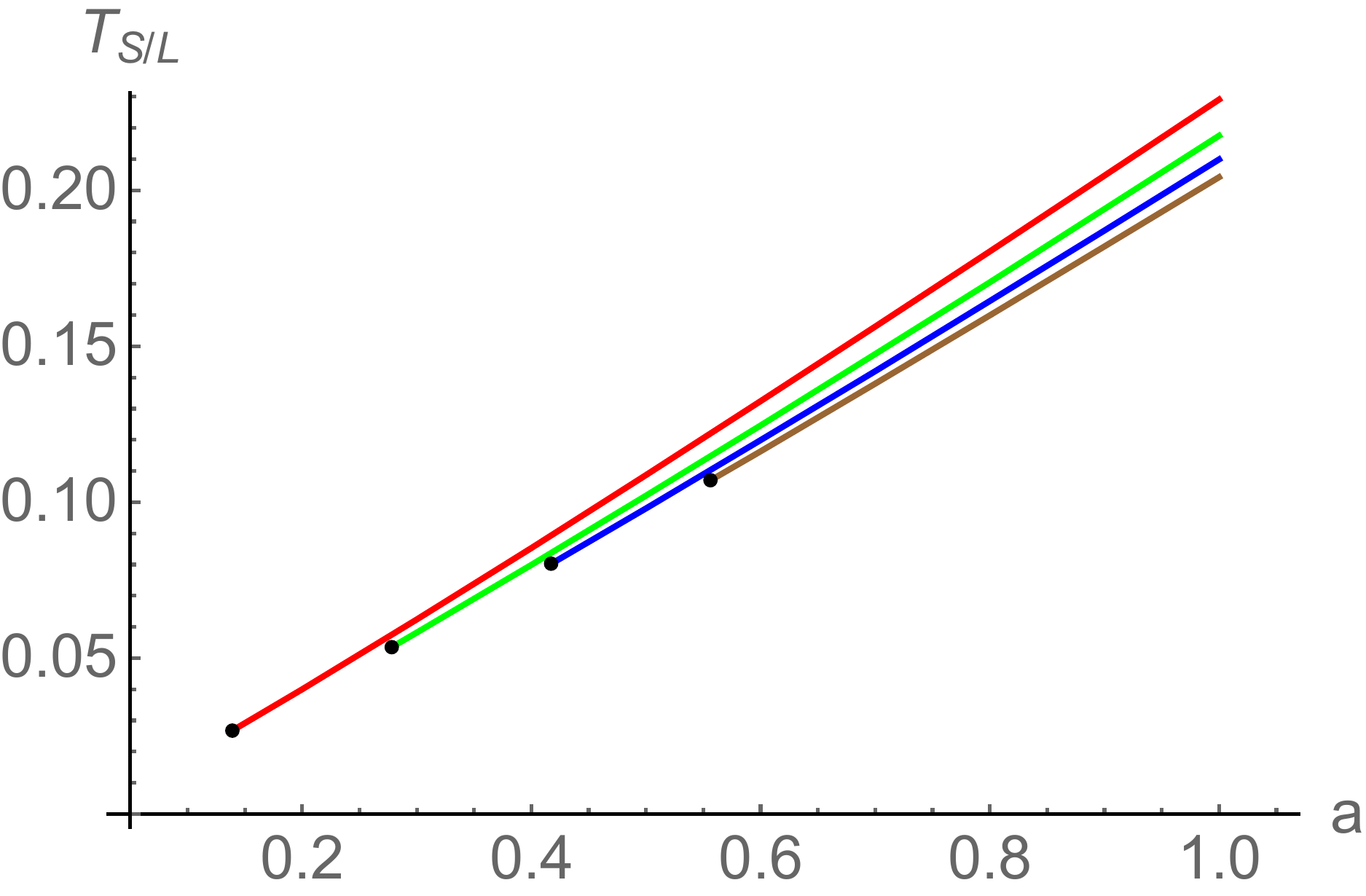}
\caption{\small The transition temperature $T_{S/L}$ as a function of $a$ for different values of $q_e$. Red, green, blue, brown, and orange curves correspond to $q_e=0.005$, $0.01$, $0.015$, and $0.02$, respectively. Black dots indicate the second-order critical points $a^{c}$.}
\label{avsTSLcritvsqef3}
\end{minipage}
\end{figure}

The thermodynamic structure of the hairy black hole is shown in Figs.~\ref{zhvsTvsaqPt01f3} and \ref{TvsFvsaqPt01f3}. The thermodynamic phase diagram for $f(\phi)=e^{-\phi^2/2}$ coupling is quite identical to the case of $f(\phi)=e^{-\phi}$ coupling. For finite $q_e$ and $a$, there are again thermodynamically stable hairy black hole phases. Here as well, the hairy charged black hole undergoes a small/large black hole phase transition and there occurs a critical charge $q_e^{c}$ at which the small/large hairy black hole phase transition line terminates and the nucleation of two stable hairy branches takes place. Similarly, there exists a critical hair strength $a_{c}$ below which the small/large hairy black hole phase transition stops. Therefore, there are again two second-order critical points $\{q_e^{c}, a_c\}$. The dependence of small/large phase transition temperature on $q_e$ and $a$ is shown in Figs.~\ref{qevsTSLcritvsaf3} and \ref{avsTSLcritvsqef3}. These figures also show how the critical points $\{q_e^{c}, a_c\}$ vary as a function of $q_e$ and $a$. These results are again quite similar to the case of $f(\phi)=e^{-\phi}$ (see Figs.~\ref{qevsTSLcritvsaf2} and \ref{avsTSLcritvsqef2}). However, there are a few differences as well. In particular, compared to $f(\phi)=e^{-\phi}$ coupling, for a fixed value of $a$, the magnitude of critical point $q_e^{c}$ is lower whereas the critical temperature $T_{S/L}^{c}$ is slightly higher. Similarly, for a fixed $q_e$, the magnitude of critical point $a_{c}$ as well the critical temperature are higher for $f(\phi)=e^{-\phi^2/2}$ coupling. Moreover, the magnitude of the extremal horizon radius also decreases substantially for $f(\phi)=e^{-\phi^2/2}$ coupling compared to $f(\phi)=e^{-\phi}$ coupling.

\section{CONSERVED MASS AND PRIMARY HAIR}
In order to show that the constructed hairy black holes are of primary nature, we need to compute the conserved charges and show that they depend only on the respective independent integration constants. We have already established in Eq.~(\ref{Qesol}) that the black hole electric charge $Q_e$ depends only on the integration constants $q_e$. To show that the conserved mass of the hairy black hole also depends on the integration constant, we need to explicitly compute the black hole mass. To illustrate this, we focus on $q_e=0$ case and rely on the holographic renormalization procedure to compute the conserved mass \cite{Balasubramanian:1999re}.  In this procedure, the conserved thermodynamic quantities are calculated from the regularized action using boundary counterterms.  For our Einstein-scalar system in
Eq.~(\ref{actionEF}), the same can be computed by subtracting the boundary terms from the bulk on shell action
\begin{eqnarray}
S_{ren} = S_{ES}^{on-shell} + \frac{1}{8 \pi G_3} \int_{\partial \mathcal{M}} \mathrm{d^2}x \ \sqrt{-\gamma} \Theta -\frac{1}{16 \pi G_3} \int_{\partial \mathcal{M}} \mathrm{d^2}x \ 2 \sqrt{-\gamma}
+ S_{b}(\phi)  \,,
\label{actionreg3D}
\end{eqnarray}
where the first term is the on-shell action, the second term is the usual Gibbons-Hawking surface term, the third term is the Balasubramanian-Kraus counterterms, and the fourth term is the scalar counterterm. The scalar counterterm is added to make sure that the scalar field has a well-defined equation of motion. In particular, the variation of the scalar kinetic term contains the boundary term
\begin{eqnarray}
\delta_\phi S_{ES} = - \frac{1}{16 \pi G_3} \int_{\partial \mathcal{M}} \mathrm{d^2}x \ \sqrt{-\gamma} n^r \partial_r \phi \delta\phi \,,
\label{variationphi3D}
\end{eqnarray}
and to satisfy the scalar equation of motion, we add a boundary term
\begin{eqnarray}
S_{b}(\phi) = \frac{1}{16 \pi G_3} \int_{\partial \mathcal{M}} \mathrm{d^2}x \ \sqrt{-\gamma} \left( \phi n^r \partial_r \phi + \frac{1}{2} \phi^2   \right)\,,
\label{boundaryphiaction3D}
\end{eqnarray}
to have a well-defined variational problem. $\gamma$ is the induced metric on the boundary $\partial \mathcal{M}$ and $\Theta$ is the trace of the extrinsic curvature $\Theta_{\mu\nu}$. From $S_{ren}$, and using the Arnowitt-Deser-Misner (ADM) decomposition, we can compute the corresponding stress energy tensor
\begin{eqnarray}
 T^{\mu\nu} =\frac{1}{8 \pi G_3} \left[ \Theta \gamma^{\mu\nu} - \Theta^{\mu\nu} + \frac{2}{\sqrt{-\gamma}}\frac{\delta\mathcal{L}_{ct}}{\delta \gamma_{\mu\nu}} \right] \,,
\label{stresstensordef3D}
\end{eqnarray}
where $\mathcal{L}_{ct}$ is the Lagrangian of the counterterms only. Explicitly, in our case, we have
\begin{eqnarray}
T_{\mu\nu} =\frac{1}{8 \pi G_3} \left[ \Theta \gamma_{\mu\nu} - \Theta_{\mu\nu}- \gamma_{\mu\nu} + \gamma_{\mu\nu} \left( \frac{\phi}{2} n^r \partial_r \phi + \frac{1}{4}\phi^2 \right)  \right]  \,.
\label{stresstensor3D}
\end{eqnarray}
The mass of the hairy black hole is then related to the $tt$ component of $T_{\mu\nu}$. In particular, if $K^\mu$ is a Killing vector generating an isometry of the boundary space, then the associated conserved charge is
\begin{eqnarray}
M = \int_\Sigma \ d x \sqrt{\sigma} u^{\mu} T_{\mu\nu} K^{\nu}  \,,
\label{massexpression3D}
\end{eqnarray}
where $\Sigma$ is a spacelike surface in $\partial\mathcal{M}$, with induced metric $\sigma$, and $u_{\mu}=-\sqrt{g(z)}\delta^{t}_{\mu}$ is the timelike unit normal to $\Sigma$.

To explicitly evaluate the mass and show that it is proportional to the integration constant, let us first write down the metric coefficient $g(z)$ in the following form
\begin{eqnarray}
g(z) = 1 + \mathcal{C} \frac{\log{(1+a^2 z^{2})}}{2 a^2} \,,
\label{gzsolc1}
\end{eqnarray}
where $\mathcal{C}$ is the integration constant coming from solving Eq.~(\ref{geom}). By requiring $g(z_h)=0$, we have its explicit form
\begin{eqnarray}
\mathcal{C} =  - \frac{2 a^2}{\log{(1+a^2 z_{h}^{2})}} \,.
\end{eqnarray}
Let us also note the near the boundary expansion of $g(z)$
\begin{eqnarray}
g(z) = 1 + \frac{\mathcal{C} z^2}{2} + \mathcal{O}(z^4) \,,
\end{eqnarray}
Substituting Eq.~(\ref{gzsolc1}) into Eq.~(\ref{massexpression3D}) and simplifying, we have the hairy black hole mass expression
\begin{eqnarray}
M = - \frac{\mathcal{C} \Omega_{1}}{32 \pi G_3}  \,,
\label{mass3D}
\end{eqnarray}
where $ \Omega_{1}=2\pi$ is the unit volume of the boundary space constant hypersurface. Notice that the mass is proportional to the constant $\mathcal{C}$, suggesting that the black hole hair is of the primary nature. Moreover, this expression also matches with the $z^2$ coefficient of $g(z)$. In particular,
\begin{eqnarray}
M = - \frac{\Omega_{1}}{16 \pi G_3} \times \left[ \text{$z^2$ coefficient of $g(z)$}  \right]\,.
\end{eqnarray}
Now, Substituting the expression of $\mathcal{C}$ into $M$, we have
\begin{eqnarray}
M = \frac{\Omega_{1}}{16 \pi G_3} \frac{ a^2}{\log{(1+a^2 z_{h}^{2})}}\,,
\end{eqnarray}
which smoothly reduces to the BTZ black hole mass expression $M=\Omega_{1}/(16 \pi z_{h}^2)$ in the limit $a\rightarrow 0$.

From the renormalized action, we can further calculate the free energy $F=-S_{ren}/\beta$
\begin{eqnarray}
& & F = \frac{\Omega_{1} \mathcal{C}}{32 \pi G_3} = - \frac{\Omega_{1}}{16 \pi G_3} \frac{ a^2}{\log{(1+a^2 z_{h}^{2})}} \,,
\label{Gibbspanar}
\end{eqnarray}
which also reduces to the BTZ free energy expression in the limit $a\rightarrow 0$. Importantly, this expression of free energy agrees with the expected thermodynamic relation $F=M - T S_{BH}$. This is a consistency check for the thermodynamic results found here for the three-dimensional hairy black holes. Moreover, we calculated the pressure, \footnote{The pressure can be computed from the $\varphi\varphi$ component of $T_{\mu\nu}$.} and find that the hairy gravity system further satisfies the standard relation,
\begin{eqnarray}
F = - P\,.
\label{pressureplanar}
\end{eqnarray}
However, unfortunately, free energy does not satisfy the differential form of the first law $F = -S_{BH}dT$. This undesirable result might be correlated to the fact that with hair this form needs to be expanded by additional terms. Indeed, many works in recent years have suggested that the differential form of the first law needs to be modified in the presence of a scalar field \cite{Liu:2013gja,Lu:2014maa}. It is of course great importance to clearly establish the first law in our hairy model; however, since our main aim in this work is on the construction and thermodynamic stability of three-dimensional hairy black holes (and on the corresponding nontrivial phase transitions), we therefore postpone this interesting problem for future work.

\section{CONCLUSIONS}
In this paper, we have analytically constructed new families of three-dimensional primary hair charged black hole solutions in the Einstein-Maxwell-scalar gravity theory. The obtained gravity solution is expressed in terms of two functions $f(\phi)$ and $A(z)$. We analyzed the gravity solution for three prominent and interesting forms of the coupling function $f(\phi)=1$, $f(\phi)=e^{-\phi/2}$, and $f(\phi)=e^{-\phi^2/2}$; and two different forms of $A(z)=-\log{(1+a^2 z^2)}$ and $A(z)=-a^2 z^2$. The parameter $a$ controls the strength of the primary scalar hair and in the limit $a\rightarrow 0$ all the hairy solution reduces to the standard nonhairy BTZ black hole solution. In each of these solutions: (i) the scalar field is regular everywhere outside the horizon and goes to zero at the asymptotic boundary; (ii) the Kretschmann and Ricci scalars are also finite everywhere outside the horizon, indicating the smooth nature of the constructed hairy black holes; and (ii) the scalar potential is bounded from above from its UV boundary value and reduces to the negative cosmological constant at the asymptotic boundary.

We then investigated the thermodynamic properties of the constructed hairy black hole solutions and found many interesting results. In particular, for $f(\phi)=1$, there exists a critical value of the hairy parameter $a_c$ above which the charged hairy black hole exhibits the Hawking/Page phase transition whereas below $a_c$ no such phase transition occurs. The corresponding transition temperature is also found to be increasing (decreasing) monotonically with $a$ ($q_e$). We further found that the specific heat is always positive in the thermodynamically favored black hole phase, thereby establishing the local stability of the hairy black holes.  The thermodynamic structure of the hairy black hole becomes even more interesting for $f(\phi)=e^{-\phi}$ and $f(\phi)=e^{-\phi^2/2}$ couplings. For a fixed $q_e$, now van der Waals type small/large black hole phase transition appears for higher values of $a$ whereas it ceases to exist at smaller values of $a$. Interestingly, there are now two second-order critical points $\{q_{e}^{c}, a_c\}$ at which the first-order small/large black hole phase transition line stops. The small/large phase transition temperature is also found to be decreasing with $q_e$. This behavior is completely analogous to the thermodynamic behavior of  the charged RN-AdS black holes in the canonical ensemble in four and higher dimensions. This is interesting considering that although BTZ black hole shares several features with their higher dimensional counterpart, however, their thermodynamic structure is vastly different. Similarly, by varying $a$, the thermodynamic structure becomes different from the varying charge scenario. In particular, now not only did the smaller values of $a$ lead to the dissolution of the small/large black hole phase transition but also the small/large phase transition temperature increases with $a$.

There are many directions to extend our work. It would be interesting to extend this work by finding its axisymmetric counterpart. We expect that, like in the BTZ black hole, the angular momentum might greatly modify the thermodynamic structure of the charged hairy black hole. It is also interesting to analyze the dynamical stability of the charged hairy black holes against various perturbations. Our initial investigations in this direction suggest that these hairy black holes are also dynamically stable under scalar field perturbations. Similarly, it would also be interesting to extend our discussion with nonlinear electrodynamic terms, as they are also known to greatly modify the thermodynamic structure, and see whether the results obtained here are general. Work in these directions is in progress.

\section*{Acknowledgments}
We would like to thank D.~Choudhuri and S.S.~Jena for careful reading of the manuscript and pointing out the necessary corrections. The work of S.~P.~is supported by Grant No. 16-6(DEC.2017)/2018(NET/CSIR) of UGC, India. The work of S.~M.~is supported by the Department of Science and Technology, Government of India under the Grant Agreement number IFA17-PH207 (INSPIRE Faculty Award).

\appendix

\section*{Appendix: BLACK HOLE THERMODYNAMICS WITH $A(z)=-a^2 z^2$}

In order to check the robustness of the results presented above for the hairy black holes, it is instructive as well as desirable to investigate the hairy black hole solution and thermodynamics for a different form of $A(z)$ as well. Here, we consider another simple form $A(z)=-a^2 z^2$. This simple form has been used to construct hairy black holes in higher dimensions \cite{Mahapatra:2020wym}, as well as used in AdS/QCD model building in five dimensions to reproduce real QCD-like properties from holography \cite{Dudal:2017max,Bohra:2019ebj}. It is therefore interesting to analyze this form of $A(z)$ in three dimensions as well. This form also makes sure that constructed geometry asymptotes to AdS at the boundary. For simplicity, here we mainly concentrate on $q_e=0$ case. Analogous results for finite $q_e$ can be straightforwardly obtained.

With $A(z)=-a^2 z^2$, most of our results for the hairy solution remain the same as in the previous case. We will, therefore, be very brief here. The solution for $\phi(z)$ and $g(z)$ reduce to
\begin{eqnarray}
& & \phi(z) = 2 a z\,, \nonumber\\
& &  g(z) = \frac{1-e^{a^2 \left(z_h^2-z^2\right)}}{1-e^{a^2 z_h^2}}\,.
\end{eqnarray}
These are again well-behaved functions. The scalar field is finite at and outside the horizon and vanishes only at the asymptotic boundary $z\rightarrow 0$. Similarly, the Kretschmann and Ricci scalars are finite everywhere outside the horizon. It suggests the existence of a well-behaved hairy black hole solution in three dimensions for $A(z)=-a^2 z^2$ as well.

\begin{figure}[h!]
\begin{minipage}[b]{0.5\linewidth}
\centering
\includegraphics[width=2.8in,height=2.3in]{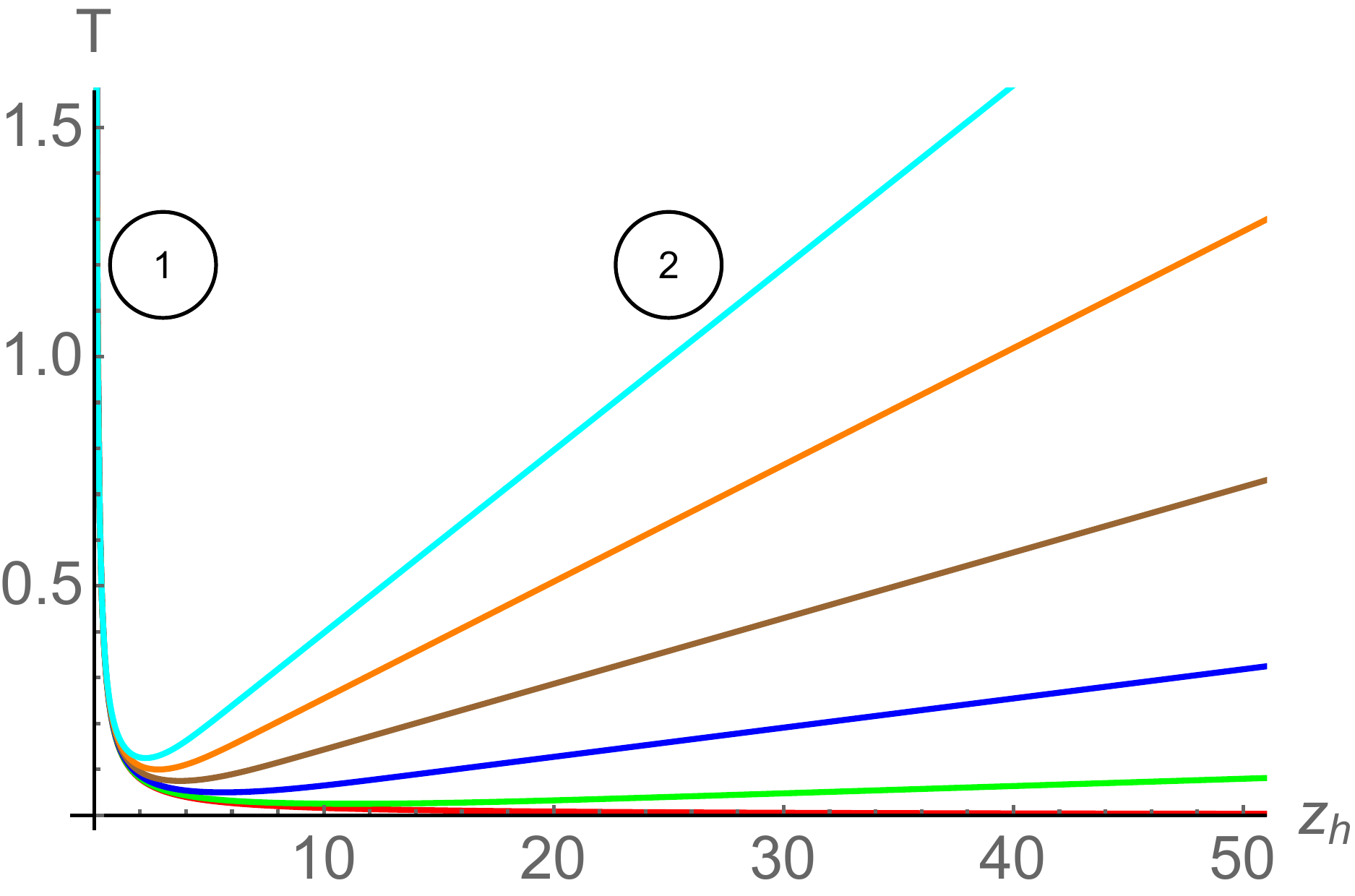}
\caption{ \small Hawking temperature $T$ as a function of horizon radius $z_h$ for various values of $a$. Red, green, blue, brown, orange, and cyan curves correspond to $a=0$, $0.1$, $0.2$, $0.3$, $0.4$, and $0.5$, respectively.}
\label{zhvsTvsaq0f1Aasq2zsq2}
\end{minipage}
\hspace{0.4cm}
\begin{minipage}[b]{0.5\linewidth}
\centering
\includegraphics[width=2.8in,height=2.3in]{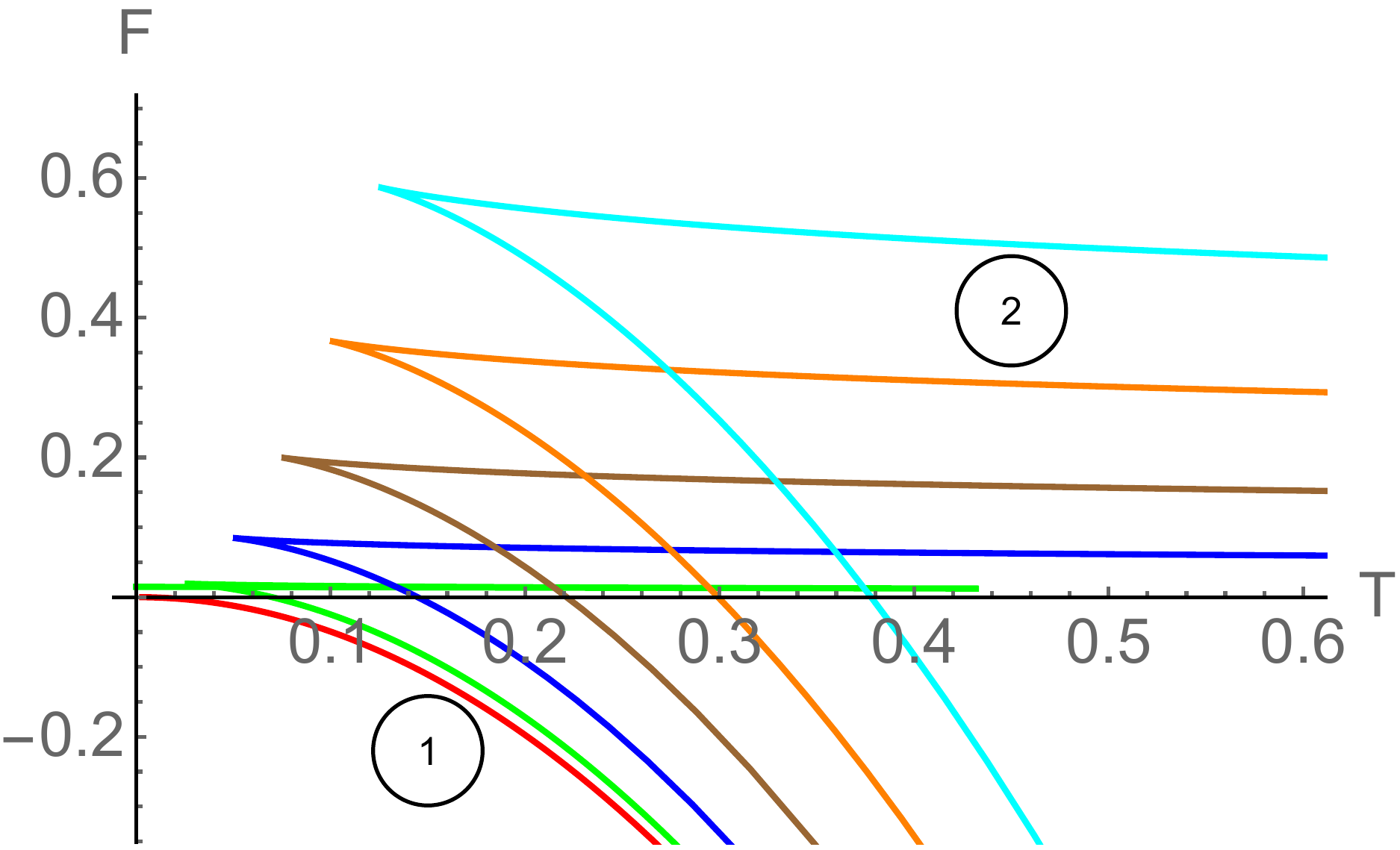}
\caption{\small Free energy $F$ as a function of Hawking temperature $T$ for various values of $a$. Red, green, blue, brown, orange, and cyan curves correspond to $a=0$, $0.1$, $0.2$, $0.3$, $0.4$, and $0.5$, respectively.}
\label{TvsFvsaq0f1Aasq2zsq2}
\end{minipage}
\end{figure}

\begin{figure}[h!]
\centering
\includegraphics[width=2.8in,height=2.3in]{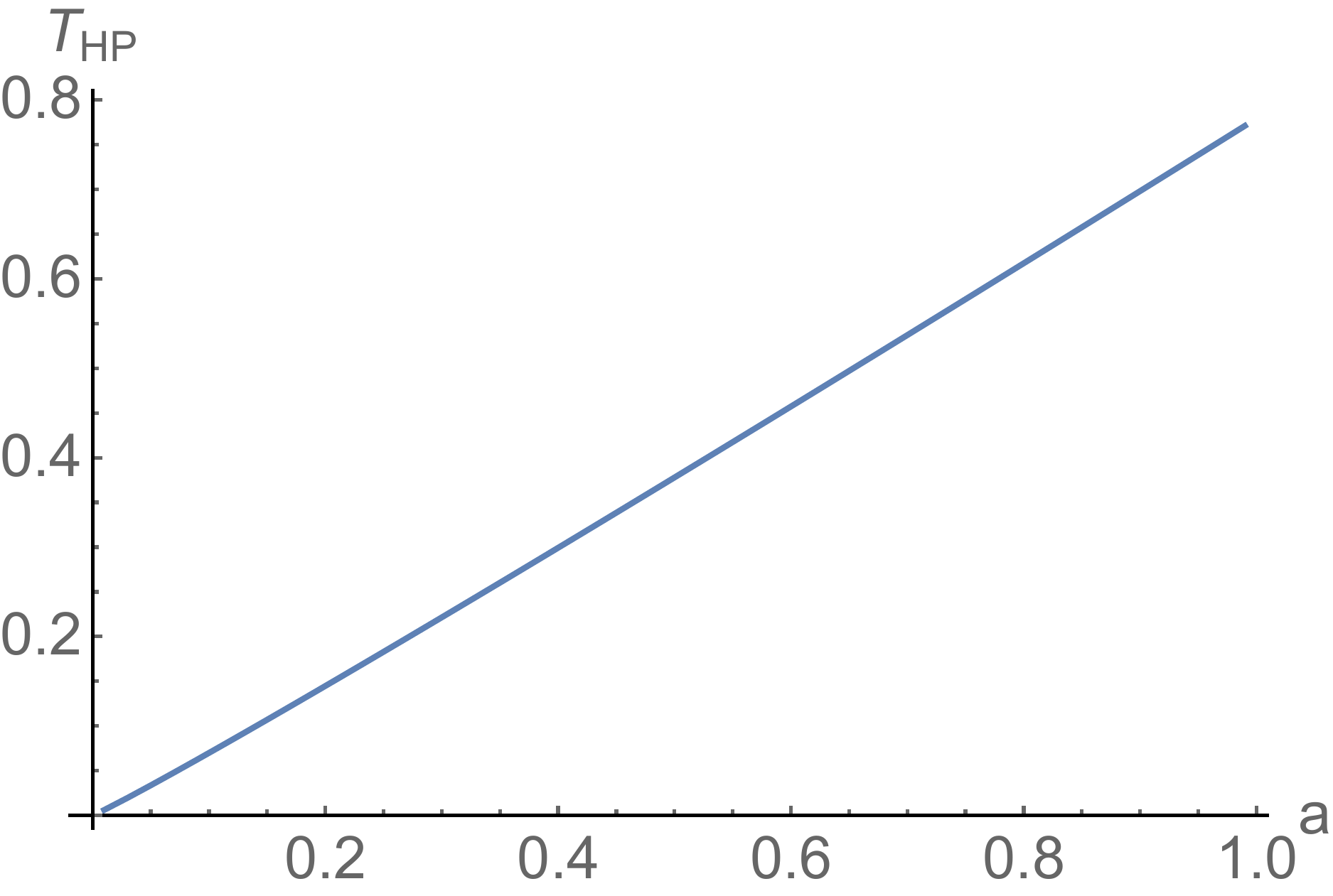}
\caption{ \small Hawking/Page phase transition temperature $T_{HP}$ as a function of $a$.}
\label{avsTHPq0f1Aasq2zsq2}
\end{figure}

We can similarly obtain the black hole temperature. It is now given by
\begin{eqnarray}
& & T = \frac{a^2 z_h e^{a^2 z_h^2}}{2 \pi  \left(e^{a^2 z_h^2}-1\right)} \,.
\label{tempAasqzsq}
\end{eqnarray}
This again has a smooth $a \rightarrow 0$ limit, reducing to the BTZ black hole expression. The temperature profiles for various values of $a$ are shown in Fig.~\ref{zhvsTvsaq0f1Aasq2zsq2}. We see that, just like for $A(z)=-\log{(1+a^2 z^2)}$, we again have two black hole branches for finite $a$. The first branch, for which the temperature decreases with $z_h$ has a positive specific heat, is stable whereas the second branch, for which temperature increases with $z_h$ has a negative specific heat, is unstable. The existence of a second unstable branch can be easily seen from Eq.~(\ref{tempAasqzsq}). In particular, for large $z_h$, $T \propto z_h$.  It also implies that there would be a minimum temperature below which no black hole branch will exist. This again suggests the possibility of Hawking /Page type phase transition between the stable black hole branch and thermal-AdS. The free energy behavior, shown in Fig.~\ref{TvsFvsaq0f1Aasq2zsq2}, confirms this fact. We can similarly compute the transition temperature. The results are shown in Fig.~\ref{avsTHPq0f1Aasq2zsq2}.  We find that, just like in the case of $A(z)=-\log{(1+a^2 z^2)}$, $T_{HP}$ varies linearly with $a$. Accordingly, the possibility for Hawking/Page phase transition again gets widened as the hair strength increases. Our overall analysis therefore again suggests the existence of thermodynamically stable hairy black hole solutions, however now, with $A(z)=-a^2 z^2$.

\bibliography{references}
\bibliographystyle{unsrt}

\end{document}